\documentclass[lettersize,journal]{IEEEtran}
\usepackage{amsmath,amsfonts}
\usepackage{array}
\usepackage[caption=false,font=normalsize,labelfont=sf,textfont=sf]{subfig}
\usepackage{textcomp}
\usepackage{stfloats}
\usepackage{url}
\usepackage{xcolor}
\usepackage{verbatim}
\usepackage{graphicx}
\usepackage{booktabs}
\usepackage{amssymb}
\usepackage{multirow}
\usepackage{makecell}
\usepackage{caption}
\usepackage{cite}
\usepackage{graphicx}
\usepackage{algorithm}
\usepackage{algpseudocode}
\usepackage[T1]{fontenc}
\usepackage[english]{babel}
\usepackage{tikz}
\newcommand{\blackcircnum}[1]{%
  \tikz[baseline=(char.base)]{
    \node[shape=circle,draw,fill=black,inner sep=1pt] (char)
    {\textcolor{white}{\textbf{#1}}};
  }%
}
\begin{document}

\title{DOPD: A Dynamic PD\text{-}Disaggregation Architecture for Maximizing Goodput in LLM Inference Serving}

\author{Junhan Liao, Minxian Xu, \textit{Senior Member, IEEE}, Wanyi Zheng, Yan Wang, \textit{Member, IEEE}, Kejiang Ye, \textit{Senior Member, IEEE}, Rajkumar Buyya, \textit{Fellow, IEEE}, Chengzhong Xu, \textit{Fellow, IEEE}
\thanks{J. Liao, M. Xu and K. Ye are with Shenzhen Institutes of Advanced Technology, Chinese Academy of Sciences, Shenzhen, China, and University of Chinese Academy of Sciences, Beijing, China.

W. Zheng is with Southern University of Science and Technology, and is also a joint-training student at the Shenzhen Institutes of Advanced Technology, Chinese Academy of Sciences, Shenzhen, China.

Yan Wang is with the College of Computer Science, Inner Mongolia University, Inner Mongolia, China.

R. Buyya is with the Quantum Cloud Computing and Distributed Systems (qCLOUDS) Laboratory, School of Computing and Information Systems, the University of Melbourne, Melbourne, Australia. 

C. Xu is with State Key Lab of IOTSC, University of Macau, Macau, China.

This work is supported by the National Key R\&D Program of China (No. 2025YFE0102017, 2025YFE0204100), National Natural Science Foundation of China under Grant 62572462, Guangdong Science and Technology Cooperation Project (No. 2025A0505020065), Guangdong Basic and Applied Basic Research Foundation (No. 2024A1515010251, 2023B1515130002), Key Research and Development and Technology Transfer Program of Inner Mongolia Autonomous Region (2025YFHH0110) and Shenzhen Science and Technology Program under Grant JCYJ20240813155810014.

M. Xu is the corresponding author.}}

\markboth{IEEE TRANSACTIONS ON SERVICES COMPUTING}%
{Shell \MakeLowercase{\textit{et al.}}: A Sample Article Using IEEEtran.cls for IEEE Journals}


\maketitle

\begin{abstract}
To meet strict Service-Level Objectives (SLO), contemporary Large Language Models (LLMs) decouple the prefill and decoding stages and place them on separate GPUs to mitigate the distinct bottlenecks inherent to each stage. However, the heterogeneity of LLM workloads causes producer-consumer imbalance between the two instance types in such disaggregated architecture.
To address this problem, we propose DOPD (Dynamic Optimal Prefill/Decoding), a dynamic LLM inference system that adjusts instance allocations to achieve an optimal prefill-to-decoding (P/D) ratio based on real-time load monitoring. Combined with an appropriate request-scheduling policy, DOPD effectively resolves imbalances between prefill and decoding instances and mitigates resource allocation mismatches due to mixed-length requests under high concurrency.
Experimental evaluations show that, compared with vLLM and DistServe (representative aggregation-based and disaggregation-based approaches), DOPD improves overall system goodput by up to $1.5\times$, decreases P90 time-to-first-token (TTFT) by up to $67.5\%$, and decreases P90 time-per-output-token (TPOT) by up to $22.8\%$. Furthermore, our dynamic P/D adjustment technique performs proactive reconfiguration based on historical load, achieving over 99\% SLO attainment while using 
fewer additional resources.
\end{abstract}

\bstctlcite{IEEEexample:BSTcontrol}
\section{Introduction}
\label{sec:Inctroduction}
\IEEEPARstart{S}{ince} Transformer-based\cite{transformer} LLMs have demonstrated powerful capabilities in natural language processing\cite{gpt-3.5}, an increasing number of researchers have pursued deeper investigations\cite{llama2, 2024qwen2} to make LLMs such as GPT-5\cite{gpt-5}, LLaMa-4\cite{Llama-4}
more capable. Concurrently, the deployment of LLMs has become increasingly pervasive across a wide range of industries and commercial applications such as advanced Bing search engine\cite{New-Bing}, Google's AI assistant \cite{gemini}, AI code editor\cite{Cursor}, log-based fault diagnosis\cite{LLM4Diagnosis}. However, more powerful models often possess extremely large parameter counts. For example, Kimi-k2 \cite{kimi-k2} contains over 1 trillion parameters. As model parameter counts grow, the computational and storage costs of inference increase substantially. Consequently, LLMs often run on expensive, energy-intensive GPUs \cite{H100}. 

Optimizing LLM inference to reduce these costs has therefore become a major research focus\cite{LLM4Edge, LLM4Mobile, BrownoutServe2025}. Both academic and industrial efforts have proposed several techniques\cite{LLM4EdgeScheduling}. For example, Key-Value (KV) caching stores historical KV tensors to avoid redundant recomputation. FlashAttention\cite{flashattention1} reduces GPU memory-access overhead when computing attention. Mixture-of-Experts (MoE)\cite{MoE} architectures substantially expand model capacity and improve performance without proportionally increasing computation.
\begin{figure}[t]
	\centering
	\includegraphics[width=\linewidth]{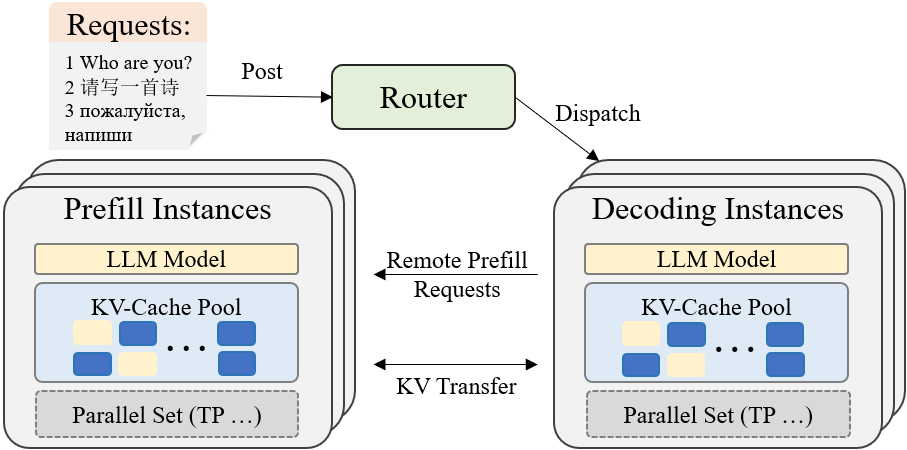}
	\caption{Prefill-decoding disaggregation diagram.}
    \label{fig:pd_summary}
\end{figure}
\textcolor{black}{Based on the unique computational characteristics of LLM inference, the process is typically divided into two stages: the prefill stage (compute-intensive) and the decoding stage (memory-intensive). To address the resource contention between these two stages, state-of-the-art systems adopt the PD-Disaggregation architecture (e.g., \text{DistServe} \cite{DistServe} and \text{SplitWise} \cite{splitwise}).  Figure~\ref{fig:pd_summary} illustrates the overall architecture of such Prefill\text{-}Decoding (PD)\text{-}Disaggregation systems, which are organized into \text{P-instances} (perform only prefill inference) and \text{D-instances} (perform only decoding inference). By decoupling the two stages of request inference onto different GPUs for processing, this architecture mitigates the interference between compute-intensive prompt processing and memory-intensive token generation. In this architecture, the P-instance acts as the producer, generating prefill-completed requests after finishing the prefill inference, while the D-instance acts as the consumer that receives these prefill-completed requests and continues with the subsequent decoding inference for each request.}

\textcolor{black}{Despite these advantages, current PD-Disaggregation methods face a critical resource optimization challenge: how to minimize resource consumption while satisfying SLO and goodput requirements. This issue is critical for organizations that provide LLM services or otherwise deploy LLMs in production, because solving it enables meeting user demand at substantially lower cost. Under current LLM workloads, a mismatch between the production capacity of P-instances and the consumption capacity of D-instances leads to severe resource inefficiency: over-provisioning results in GPU idleness and resource wastage, while under-provisioning leads to SLO violations and degraded user experience. Therefore, identifying P-instance and D-instance configurations that achieve a producer-consumer balance is the key to minimizing resource usage. Meanwhile, the heterogeneity of LLM workloads causes the optimal configuration to change over time and static deployments remain susceptible to producer-consumer imbalance. Consequently, to accommodate diverse workloads, dynamic and rationalized adjustments are essential. However, excessively frequent reconfiguring of resource allocation incurs substantial time costs, therefore mitigating the short-term mismatch between static PD-Disaggregation deployments and mixed-length request traffic remains a challenging problem.}

To address these challenges, we tackle the problem of dynamically adjusting the P/D ratio in a PD-Disaggregation deployment to improve goodput. Concretely, we (i) derive a principled method to compute an optimal P/D ratio for a given load regime, (ii) apply intelligent scheduling to mitigate the short-term resource mismatch caused by mixed-length requests and improve the effectiveness of systems in high-load scenarios, and (iii) leverage the monitoring data to re-evaluate the system P/D ratio and decide whether the current configuration remains optimal or requires elastic scaling of P-instances and D-instances. Our contributions are as follows:
\begin{itemize}
\item We design DOPD, an intelligent and dynamic LLM inference framework that continuously tunes the number and configuration of P-instances and D-instances.
\item We present a comprehensive system model and in-depth analysis of PD-Disaggregation, identifying the key challenges faced by disaggregated inference architectures.
\item We propose an optimal P/D ratio calculation method, and we design an intelligent, length-aware request-scheduling algorithm to mitigate resource mismatch caused by mixed-length inference workloads. 
\item We perform extensive experiments demonstrating that DOPD substantially increases system goodput ($1.5\times$) and the SLO attainment (from 80.8\% to 99.4\%) under realistic production traces, while validating its effectiveness across diverse scenarios. 
\end{itemize}

\begin{figure}[t]
	\centering
	\includegraphics[width=\linewidth]{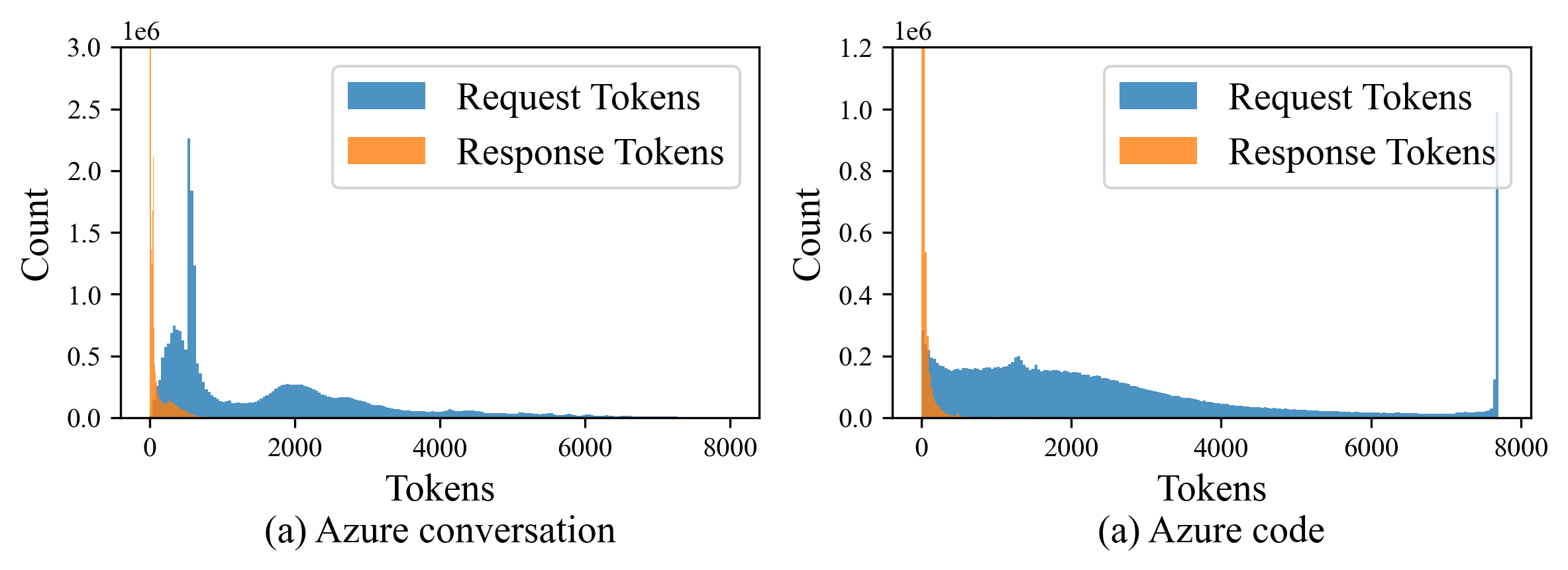}
	\caption{The input/output sequence length of request tokens in Microsoft Azure conversation and code traces.}
	\label{fig:workload_length}
\end{figure}
\section{Motivation}
\label{sec:motivation}
This section summarizes the issues we observe in existing PD-Disaggregation systems and motivates our design choices.
\subsection{Motivation: Difficulty of Predicting Inference Workloads}
Modern LLM inference workloads are highly heterogeneous and nonstationary. Individual user requests exhibit large variability in input and output sequence lengths, which makes accurate load prediction challenging. Figure~\ref{fig:workload_length} illustrates the input/output sequence length distributions from a production trace (Microsoft Azure conversation and code traces)\footnote{https://github.com/Azure/AzurePublicDataset}. These traces were collected from real user interactions with LLM services and exhibit wide length variance, lack clear periodic patterns, and frequently contain sudden bursts. Such characteristics render many traditional prediction methods ineffective, yet accurate short-term forecasts of inference load are essential for prudent resource pre-allocation. For LLM deployments, knowing the near-term inference load in advance enables proactive adjustment of instance allocations, which is critical for preventing resource waste and avoiding overload in PD-Disaggregation systems.

To mitigate this challenge, applying a multiplicative scaling (correction) factor can be effective. By comparing the most recent forecast with observed measurements and adjusting future forecasts accordingly, the scaling factor attenuates systematic bias between predicted and actual values and keeps prediction error within a bounded, small range.
\subsection{Motivation: Bad P/D Ratio Waste Resources}
\begin{figure}
    \centering
    \includegraphics[width=\linewidth]{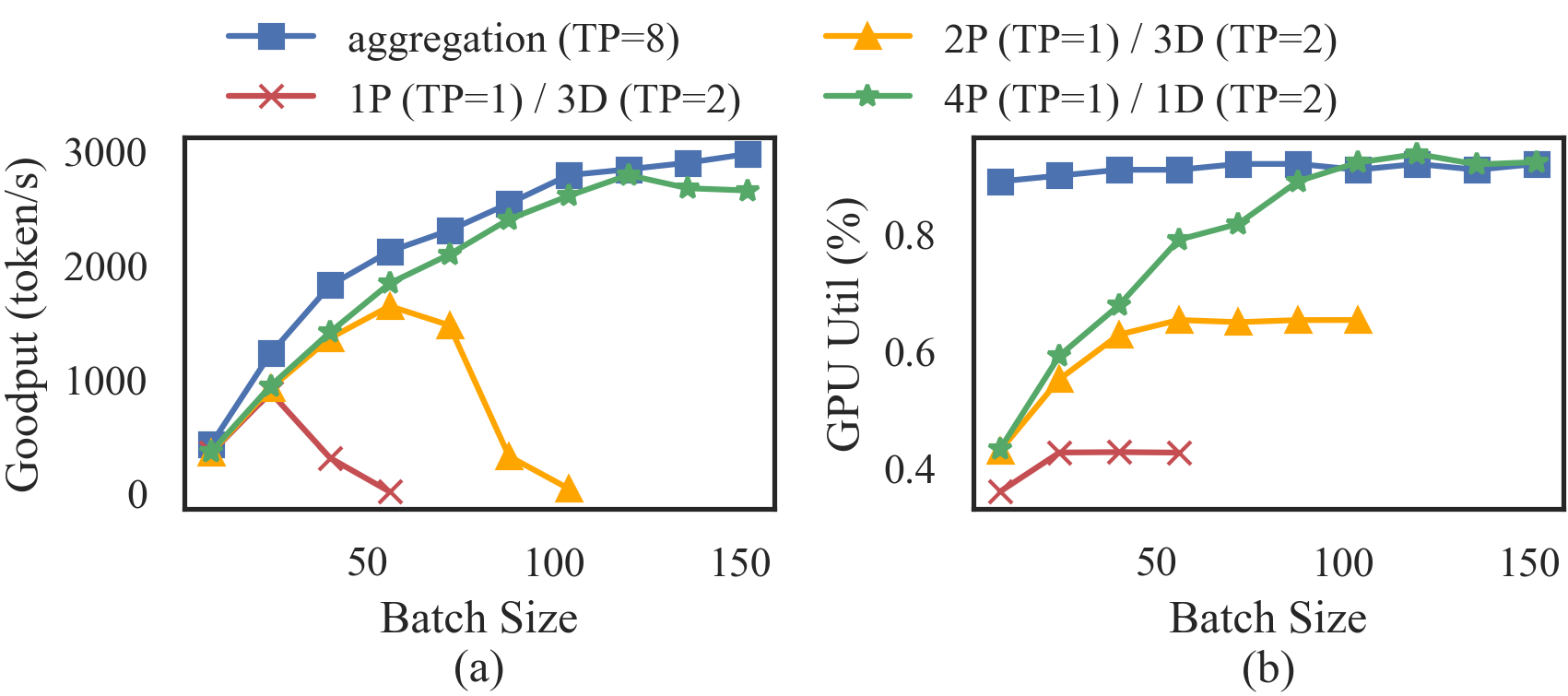}
    \caption{\textcolor{black}{The goodput (a) and GPU utilization (b) comparison with different configurations. The \text{"2P (TP=1) / 3D (TP=2)"} represents a PD-Disaggregation deployment comprising 2 P-instances each deployed on a single GPU (tensor parallel size $=1$) and 3 D-instances each deployed on two GPUs (tensor parallel size $=2$). The \text{"aggregation (TP=8)"} denotes an aggregation-based system deployed on 8 GPUs (tensor parallel size $=8$).}}
    \label{fig:motivation_diff_pd}
\end{figure}

\textcolor{black}{Under the PD-Disaggregation architecture, a mismatch between the system P/D ratio and the current workload leads to severe resource wastage, thereby causing SLO violations and significantly impairing system goodput performance.
Figure~\ref{fig:motivation_diff_pd} illustrates the direct impact of different P/D configurations on system performance. Experimental results indicate that different P/D ratios induce significant performance disparities: A PD-Disaggregation system configured with an optimal P/D ratio (the green curve) achieves goodput comparable to a collocated system deployed with 8 GPUs (the blue curve) while utilizing only 6 GPUs. Conversely, a suboptimal P/D ratio (the orange curve), despite utilizing 8 GPUs, fails to leverage the inherent advantages of PD-Disaggregation, resulting in a drastic decline in the utilization of surplus GPUs and consequently yielding poor performance. Evidently, an improper P/D ratio not only significantly deteriorates goodput and GPU utilization but also wastes limited GPU resources, thereby necessitating the system to allocate more resources to satisfy the same SLOs.}

\textcolor{black}{To mitigate the producer-consumer imbalance, we model resource allocation as an optimization problem. Given short-term workload statistics, our objective is to minimize the total number of provisioned GPUs while satisfying stringent SLO requirements for TTFT, TPOT, and goodput, subject to available hardware constraints. By framing the selection of P-instances and D-instances in this manner, we ensure cost-efficiency without compromising service quality. The specific algorithmic implementation and solver details are further elaborated in Section~\ref{subsec:optimal_pd}.}

\subsection{Motivation: Mixed-length Requests Undermine PD-Disaggregation}
Prior work \cite{variable_length, ShuffleInfer} and our observations show that when the workload is a mixture of requests with widely varying lengths, PD-Disaggregation can perform substantially worse than deployments in which all requests share a fixed length. The root cause is that different request lengths have different suitable allocations between prefill and decoding resources. Short requests tend to complete decoding quickly while consuming relatively little prefill capacity, whereas long requests require a higher proportion of prefill resources to sustain pipeline efficiency. If requests with very different lengths (e.g., 100 tokens and 1900 tokens) are mixed and a single global P/D ratio is chosen based on the average length (e.g., 1000 tokens), the resulting static allocation typically cannot satisfy the local optima for both short and long requests simultaneously. Therefore, static configurations based on average length of requests therefore produce either resource shortages or long queuing for subsets of requests, degrading overall throughput and tail latency and eroding the benefits of PD-Disaggregation.

At the same time, for extremely short requests whose prefill compute time is negligible, the KV-cache transfer latency between P-instances and D-instances becomes a nontrivial fraction of end-to-end latency. Table~\ref{tab:inference_time} reports, for a single H100 GPU\cite{H100} running an LLaMa-3.3-70B-FP8 \cite{DeepSeek-R1-Distill-Llama-70B-FP8-dynamic} model, the per-request prefill time and the per-step decoding time (under maximum batching) for several sequence lengths. For example, a prefill for a prompt of length 100 tokens may take approximately $36$ms, while a single decoding step at 248 batch size can take approximately $47$ms. Even with NIXL's\cite{nixl} efficient GPU-to-GPU transfers, KV-cache transmission can add on the order of $15$-$80$ms depending on block fragmentation and count, rendering PD-Disaggregation inefficient for such requests. In such cases, executing these ultra-short requests in-place (i.e., without remote prefill) typically produces negligible interference with other decoding work, whereas routing them through remote prefill incurs unnecessary KV-cache transfer overhead and increases TTFT. Hence, offloading all ultra-short requests to remote P-instances is often suboptimal.

These two observations motivate the need for length-aware scheduling and workload-aware adaptation. Designing scheduling policies that account for request-length heterogeneity is essential to improve the robustness and efficiency of practical inference systems under mixed workloads.

Motivated by the foregoing, we propose \textbf{DOPD}, a PD-Disaggregation framework that (i) produces accurate short-term forecasts of future load, (ii) computes the optimal P/D ratio for the current workload, (iii) leverages workload-aware request scheduling to mitigate the interference introduced by mixed-length requests in PD-Disaggregation under high concurrency, and (iv) dynamically scales the numbers of P-instances and D-instances according to the computed optimal P/D ratio. By reacting to complex and rapidly varying loads, DOPD maximizes goodput using fewer GPU resources, reduces TPOT and end-to-end latency, and maintains excellent SLO attainment.
\begin{table}[t]
  \centering
  \caption{For different sequence lengths, we measure the single-run prefill inference latency and the single-step decoding inference latency at different concurrency, where cc refers to request concurrency in the system.}
  \label{tab:inference_time}
  \begin{tabular}{l|c|ccr} 
    \toprule
    \multirow{2}{*}{Sequence Length} 
      & \multirow{2}{*}{$T_{p}$ (ms)} 
      & \multicolumn{3}{c}{$T_{d}$ (ms)} \\ 
    \cline{3-5} 
     & & $cc=104$ & $cc=200$ & $cc=248$\\
    \midrule
    100 & 36 & 28 & 45  & 47 \\
    200 & 46 & 31 & 46  & 49 \\
    700 & 125 & 33 & 47  & 53 \\
    1200 & 193 & 35 & 50  & 57 \\
    1700 & 269 & 37 & 53  & 61 \\
    \bottomrule
  \end{tabular}
\end{table}

\section{Related Work}
\label{sec:related}
This section surveys recent techniques that mitigate interference between prefill and decoding requests and discusses their limitations.

\subsection{Aggregation-based Approaches}
Yu et~al. propose Orca \cite{Orca}, which uses continuous batching (iteration-level batching and selective batching) applied to generation to mitigate interference from prefill requests within the same batch. Kwon et~al. introduce vLLM \cite{vllm}, which separates prefill and decoding requests into different batches and provides optimized operators for the two request types to improve computational efficiency. Both approaches partially alleviate prefill-to-decoding interference, but long-context prefill requests can still delay fast decoding requests.

Agrawal et~al. present Sarathi-Serve \cite{taming}, introducing chunked prefill and continuous batching to split long prefill work into smaller chunks that can be interleaved with decoding work. Building on mixed-batch techniques, Kamath et~al. propose POD-Attention \cite{PODAttention}, a kernel-level redesign of the attention operator that enables true prefill–decoding overlap on a single GPU by partitioning computation and memory resources at the streaming-multiprocessor level. In practice, POD-Attention further mitigates low decoding throughput and prefill resource domination, yet these methods exhibit stability limitations under high concurrency, long-context workloads, or strict TTFT requirements.

Additionally, some other works propose reducing prefill-stage computation to mitigate the interference caused by co-processing prefill and decoding requests, for example RadixAttention \cite{sglang} and Native Sparse Attention \cite{NSA}. However these approaches are not universally applicable and some may introduce accuracy degradation.

In summary, aggregation-based approaches, such as continuous batching, chunked prefill, kernel-level attention redesign, and sparse attention techniques yield tangible gains but rely on complex, hardware-dependent kernel implementations, introduce additional implementation or memory overheads, and may compromise accuracy in some scenarios. More importantly, they do not fundamentally eliminate the conflict between prefill and decoding requests. By contrast, DOPD employs targeted, length-aware request scheduling to selectively separate or aggregate requests, achieving stronger SLO attainment while preserving high goodput.

\subsection{Disaggregation-based Approaches}
Zhong et~al. formalize the cost of colocated execution in DistServe \cite{DistServe}, demonstrating that colocating prefill and decoding causes strong interference and proposing fully disaggregated deployments to improve goodput. 
Patel et~al. build on PD-Disaggregation to present SplitWise \cite{splitwise}, which maximizes computation–communication overlap via a hierarchical KV-cache transfer design and thereby reduces the impact of KV-cache transfer overhead. Similarly, Qin et~al. propose Mooncake \cite{Mooncake}, a KV-cache–centric disaggregated architecture that optimizes prefill-to-decoding KV-cache migration via D2D/RDMA transport and KV-aware scheduling. 
However, these methods are largely static and do not support real-time dynamic adjustment of instance counts, which can lead to substantial resource waste and transient overloads under variable workloads.
\begin{table}[t]
  \centering
  \caption{Comparison of related work which is disaggregation-based Approach.}
  \label{tab:relatedwork_comparison}
  \begin{tabular}{c|cccc} 
    \toprule
    \multirow{2}{*}{\textbf{Approaches}} & \multirow{2}{*}{\makecell{\textbf{Dynamic} \\ \textbf{P/D}}} & \multirow{2}{*}{\makecell{\textbf{Efficient} \\ \textbf{KV-cache Transfer}}} & \multirow{2}{*}{\makecell{\textbf{Optimal} \\ \textbf{P/D}}} \\
    \\
    \midrule
    Zhong et al. \cite{DistServe} & $\times$ & $\times$ & $\times$ \\
    Patel et al. \cite{splitwise} & $\times$ & \checkmark & $\times$ \\
    Qin et al. \cite{Mooncake} & $\times$ & \checkmark & $\times$ \\
    Jin et al. \cite{pdserve} & \checkmark & \checkmark & $\times$ \\
    Wu et al. \cite{arrow} & \checkmark & \checkmark & $\times$ \\
    Dynamo \cite{ai-dynamo} & \checkmark & \checkmark & $\times$ \\
    \textbf{Our Work (DOPD)} & \textbf{\checkmark} & \textbf{\checkmark} & \textbf{\checkmark} \\
    \bottomrule
  \end{tabular}
\end{table}

Jin et~al. propose PD-Serve \cite{pdserve}, which constructs fine-grained P/D grouping by scenario so that requests with similar prefixes are processed within the same group, and which optimizes KV-cache transfer and encoding parallelism while supporting dynamic scaling. 
Likewise, Wu et~al. propose Arrow \cite{arrow}, which designs adaptive scheduling and an elastic instance pool that dynamically adjusts P-instance and D-instance counts based on cluster metrics, improving robustness to bursts and fluctuations. More recently, NVIDIA released the Dynamo framework \cite{ai-dynamo}, which advocates prefill–decoding separation to enable tailored tensor parallel size and memory strategies and parallelism. 
Although these systems introduce dynamic expansion, they still fall short in selecting optimal P/D instance counts and configurations after scaling: achieving post-scale optimal P/D configurations remains difficult, often resulting in significant GPU idleness or overload.

A comparison of representative characteristics of prominent PD-Disaggregation studies is presented in Table~\ref{tab:relatedwork_comparison}. In summary, the disaggregation-based of approaches decouple prefill and decoding and, on that basis, optimize KV-cache migration and transfer to overlap computation and communication (e.g., SplitWise's hierarchical transfer), and introduce fine-grained grouping and elastic/adaptive scheduling (e.g., PD-Serve's P/D grouping). These measures reduce direct resource contention between the two request types.
But many early solutions lack support for real-time, fine-grained elastic scaling, causing resource waste or short-term overloads under fluctuating load. And even systems that support dynamic scaling still face the unresolved challenge of accurately selecting the optimal number and configuration of P/D instances after scaling to avoid GPU idleness or overload, in addition to the engineering cost of complex schedulers and KV-cache migrations.

The method proposed in this paper addresses these gaps by forecasting near-term load and proactively scaling P-instances and D-instances according to an analytically derived optimal P/D ratio, thereby mitigating resource waste and overload caused by suboptimal P/D instance scheduling.
\section{System Model}\label{sec:modeling}
This section presents the system model of DOPD, covering the following aspects: 1) Workload characterization. 2) The optimal P/D ratio calculation. 3) Modeling of P-instance. 4) Modeling of D-instance.
\subsection{Workload Characterization}
Unlike traditional machine-learning workloads with fixed input dimensionality, LLMs workloads are substantially more heterogeneous. Each user request to an LLM may have a different input sequence length, which in turn yields different per-request computational costs. Because mainstream LLMs are Transformer-based, the computational pressure concentrates in the attention and feed-forward components. The attention computation can be written as 
\begin{equation}
    Output_{Attn} = \operatorname{softmax}(QK^T)V,
    \label{eq:Attention}
\end{equation} 
where $Q,K,V$ are the query, key, and value matrices produced from the input sequence by linear projections. The asymptotic complexity of the attention operation is 
\begin{equation}
    C_{attn}(ISL)=\mathcal{O}\big(ISL\cdot H^2 + ISL^2\cdot H\big),
    \label{eq:attn_cost}
\end{equation}
where $ISL$ denotes the input sequence length and $H$ denotes the model hidden size. The feed-forward computation can be abstracted as 
\begin{equation}
    Output_{FFN} = Linear(Linear(Linear(Output_{Attn}))),
    \label{eq:FFN}
\end{equation}
whose complexity is approximately
\begin{equation}
    C_{ffn}(ISL)=\mathcal{O}\big(ISL\cdot H\cdot H^{'}\big),
    \label{eq:ffn_cost}
\end{equation}
where $H^{'}$ denotes the intermediate size of the feed-forward layer (typically $H^{'}\ge H$). Thus the per-request computational cost can be expressed as
\begin{equation}
    C(ISL)=C_{attn}(ISL)+C_{ffn}(ISL).
    \label{eq:total_cost}
\end{equation}

To reason about heterogeneous requests we lift this per-request cost to the level of the input sequence length distribution $f_{ISL}$ and consider moments of $ISL$. In particular the expected per-request cost is
\begin{equation}
\begin{split}
    \mathbb{E}[C]&=\int C(x)\,f_{ISL}(x)\,dx \\
    &= \kappa_1 H^2\mathbb{E}[ISL]+\kappa_2 H\mathbb{E}[ISL^2]+\kappa_3 HH^{'}\,\mathbb{E}[ISL],
    \label{eq:expected_cost}
\end{split}
\end{equation}
where constants $\kappa_1,\kappa_2,\kappa_3$ hide implementation-level factors. Note that
\begin{equation}
    \mathbb{E}[ISL^2]=\operatorname{Var}(ISL)+\mathbb{E}[ISL]^2,
    \label{eq:second_moment}
\end{equation}
and because of the $ISL^2$ term in Eq.~\eqref{eq:attn_cost}, the cost function $C(ISL)$ is convex in $ISL$, so by Jensen's inequality we have
\begin{equation}
    \mathbb{E}[C]\ge C\big(\mathbb{E}[ISL]\big),
    \label{eq:jensen}
\end{equation}
which quantifies how heterogeneity (and in particular variance) amplifies average computation beyond what a mean-length estimate would predict.

Empirically, inference workloads commonly exhibit a long-tail distribution: a small fraction of extremely long requests requires the service to retain the capability to handle such tails. A standard heavy-tail model is the Pareto law
\begin{equation}
    \Pr\{ISL>x\}=\Big(\frac{x_m}{x}\Big)^{\alpha}, \; s.t.\;  x\ge x_m,\ \alpha>0,
    \label{eq:pareto}
\end{equation}
where the tail index $\alpha$ controls the existence of moments (for example, the variance is finite only if $\alpha>2$). When $\alpha$ is small the second moment in (Eq.~\eqref{eq:second_moment}) can be very large (or formally divergent), explaining why a tiny fraction of requests can dominate overall compute and memory pressure.

The high variance in input length makes aggregating prefill and decoding requests problematic. Because the per-step inference time of the two request types can differ dramatically, batching them together causes severe mutual interference and substantially degrades system performance. This effect can be formalized through queueing approximations. If requests arrive with (time-varying) rate $\lambda(t)$ and service time $S$ satisfies $\mathbb{E}[S]\propto \mathbb{E}[C]/COMPUTE_{speed}$, then Little's law $L=\lambda W$ links occupancy $L$ and mean sojourn $W$, and Kingman's heavy-traffic approximation for a $G/G/1$ queue gives the mean waiting time
\begin{equation}
    W_q\approx \frac{\rho}{1-\rho}\cdot\frac{c_a^2+c_s^2}{2}\,\mathbb{E}[S],
    \label{eq:kingman}
\end{equation}
where $\rho=\lambda \mathbb{E}[S]$ is load, $c_a^2$ is the squared coefficient of variation of inter-arrival times and $c_s^2=\operatorname{Var}(S)/\mathbb{E}[S]^2$ is that of service time. Because $c_s^2$ grows with $\operatorname{Var}(ISL)$ via (Eq.~\eqref{eq:total_cost}) and (Eq.~\eqref{eq:second_moment}), mixing requests with very different $ISL$ (or mixing prefill and decoding) inflates $c_s^2$, increasing $W_q$ and thus end-to-end latency.

The complexity of LLM inference service workloads extends further. Requests arrive in short-lived bursts and the instantaneous concurrency seen by the service varies over time. Such burstiness can be modeled by Markov-Modulated Poisson Processes (MMPP) or compound-Poisson models, which capture non-stationary and over-dispersed arrivals (i.e., $c_a^2\!>\!1$). The peak concurrency in the next service epoch may be orders of magnitude larger (or smaller) than in the previous epoch. Consequently, adaptive resource provisioning that accounts for both the distribution of request lengths (via moments such as $\mathbb{E}[ISL],\ \operatorname{Var}(ISL)$ and tail index $\alpha$) and rapidly changing concurrency (via time-varying $\lambda(t)$ or MMPP parameters) is essential for maximizing resource utilization. This non-stationary and bursty nature can cause serious resource waste in PD-Disaggregation systems because a fixed P/D ratio is suitable only for a restricted class of workloads.

To address these challenges, DOPD periodically feeds background-collected telemetry into an ARIMA-based\cite{ARIMA} load predictor and issues regular forecasts of the future average request length and concurrency. These forecasts are combined with the moment-based cost estimates in Eq.~\eqref{eq:expected_cost} and the queueing-delay estimates such as Eq.~\eqref{eq:kingman} to drive DOPD's dynamic P/D ratio computation and elastic resizing of P-instances and D-instances, enabling the system to pre-allocate resources commensurate with the anticipated heterogeneous and time-varying workload.
\subsection{The Optimal P/D Ratio Calculation}\label{subsec:optimal_pd}
\subsubsection{The Object of this Optimal P/D Ratio}
When all requests are served via remote prefill (perform prefill in P-instance), the prefill-to-decoding workflow naturally forms a producer-consumer pair. P-instances produce prefill-completed requests (together with KV-cache), and D-instances consume those prefill-completed requests to perform token generation. Therefore, balancing production and consumption is essential to reduce resource waste.
And system throughput for disaggregated inference is largely determined by the concurrency (the number of in-flight requests in system) that each D-instance can sustain, because each generated token must be produced by a D-instance and returned to the frontend. The maximum concurrency a D-instance can support depends on request length as well as on GPU memory capacity and memory bandwidth. Therefore, the optimal P/D ratio should (i) maximize the aggregate concurrency supported by D-instances and (ii) select a corresponding number of P-instances so that minimize idle time for P-instances and D-instances.

\subsubsection{Method for Computing the Optimal P/D Ratio}
To simplify the following calculations, assume all incoming requests have the same input sequence length $ISL$ and output sequence length $OSL$. First, based on GPU memory bandwidth $BW$, GPU memory capacity $\mathit{M}_{gpu}$, model size $\mathit{M}_{model}$, and the CUDA graph inference configuration in use, determine a suitable parallelization configuration (e.g., tensor parallelism). Note that the maximum parallelism of a D-instance must be bounded by both memory capacity and memory bandwidth constraints. 
From the memory-apacity perspective, for tensor parallel size $TP$, the remaining memory available for KV-cache on a D-instance is
\begin{equation}
    V_{mem} = (M_{gpu} - M^{'})\times TP-M_{model},
    \label{eq:memory_capacity}
\end{equation}
where $M^{'}$ encompasses peak activation‐parameter memory (which varies with the model's maximum sequence length) and other ancillary GPU memory usage.

\textcolor{black}{From the perspective of memory bandwidth, as the GPU must continuously retrieve the KV-cache from its own High Bandwidth Memory (HBM) during inference, and given that the computational load for each request during the decoding stage is minimal, the vast majority of execution time is spent accessing model weights and KV-cache. Consequently, the constrained memory bandwidth hinders the retrieval of model weights and KV-cache for all concurrent requests within the TPOT interval specified by the SLO, leading to inevitable SLO violations. To conservatively estimate the maximum concurrency under memory bandwidth limitations, we can neglect the time spent on computation and assume that the entire decoding inference latency for a request under high concurrency is dedicated to reading the KV-cache and model weights. 
So, we can calculate the maximum volume of KV-cache data that a D-instance can retrieve from HBM within the time interval $t_{d}^{SLO}$ (the TPOT SLO limit) as followed:}
\begin{equation}
    V_{BW}=t_{d}^{SLO}\times \alpha \times TP\times BW,
    \label{eq:memory_bandwidth}
\end{equation}
\textcolor{black}{Based on this data volume, we can calculate how many requests' KV-cache can be retrieved from HBM by a single D-instance within $t_{d}^{SLO}$. Consequently, by taking the minimum of the KV-cache data volume storable on the D-instance and the total KV-cache data volume retrievable within a single decoding step, and dividing this value by the memory footprint of a single request's KV-cache, we can derive the maximum concurrency $cc_d$ of the D-instance under both memory capacity and memory bandwidth constraints.}
\begin{equation}
    cc_d = \frac{\min(V_{mem}, V_{BW})}{(ISL+OSL/2)\times M_{token}},
    \label{eq:concurrency_decoding}
\end{equation}

In practice, to improve headroom for bursty traffic one may choose $TP$ that an instance can support up to $0.9\times cc_g$ concurrency. 
After obtaining the D‐instance's suitable $TP$ and maximum concurrency, we calculate the required production capacity of a single P-instance. Since the prefill stage often becomes the compute bottleneck, some systems set the P‐instance batch size as low as 1. 
\textcolor{black}{For simplicity of calculation, here we assume that each P-instance inference only generates one prefill-completed request. Let $t_p$ denote the time for a P-instance to perform a single prefill operation, and $t_d$ denote the time for a D-instance to perform a single decoding step at its maximum concurrency. Within the time interval required for a D-instance to complete the full decoding sequence of a request (which is $t_d \times OSL$), the P-instance generates a new prefill-completed request every $t_p$. Each time the D-instance receives a prefill-completed request, its concurrency increases by one. Therefore, the maximum concurrency that a single P-instance can drive on a D-instance is equal to $\frac{t_d \times OSL}{t_p}$. Since our objective for the optimal P/D ratio is to ensure that every D-instance reaches its maximum concurrency $cc_d$, the optimal P/D ratio satisfies:}
\begin{equation}
    n_p \times \frac{t_d\times OSL}{t_p} = n_d \times cc_{d},
    \label{eq:npnd}
\end{equation}
Thus, we derive an analytical formula for the optimal P/D ratio. 

Applying this method yields a deployment configuration that, for a given load, selects the resource allocation between P-instances and D-instances so that both roles exhibit near-zero idle time during steady-state inference and the overall system throughput is maximized. The proposed DOPD system employs an ARIMA-based predictor to forecast near-term workload characteristics from historical telemetry, including the future average $ISL$, the future average $OSL$, and the expected average concurrency. DOPD maps these forecasts to profiled Prefill and Decode performance metrics, computes the optimal P/D ratio and the corresponding instance configurations, and then enacts non-disruptive instance adjustments. The detailed calculation and execution steps are described in subsection~\ref{subsec:pd_manager} and Algorithm~\ref{alg:pd_adjust}.

\subsection{Modeling of  P-instance}
\textbf{For P-instances, an excessively large $TP$ (tensor parallel size) can be detrimental.} Prefill requests are compute-bound, which imposes high demands on GPU compute capability. \textcolor{black}{Using tensor parallelism allows parallel execution of the same request across multiple GPUs, thereby distributing the computational load and providing acceleration. However, tensor parallelism also introduces two All-Reduce operations per Transformer layer during inference communication-heavy operations that synchronize partial results across GPUs.} To model this trade-off briefly, let $C(ISL)$) denote the compute work for a prefill request of input length $ISL$. We formulate the prefill latency under a given $TP$ as the sum of an ideally accelerated compute term and a communication term:

\begin{equation}
\label{eq:prefill_total}
T_{p}(ISL,TP)
\;=\;\frac{C(ISL)}{S(TP)} \;+\; \tau(TP),
\end{equation}
where $S(TP)$ is the effective speedup (with $S(TP)\le TP$ and typically showing diminishing returns) and $\tau(TP)$ models tensor parallelism-induced communication overhead (which is non-decreasing in $TP$). 
\begin{algorithm}[t]
\caption{Prefill Requests Scheduler}
\label{alg:prefill_request_scheduler}
\begin{algorithmic}[1]
\State \textbf{Input:}
\State $i_{next}$: Length threshold for immediate processing
\State \textit{queue}: Remote prefill requests queue 
\State \textit{timer}: Record request wait time
\State \textit{batch}: Collection of requests to be batched
\State \textbf{Initialize} \textit{timer}, \textit{batch}, \textit{queue}
\Loop
  \State Set waiting timeout for \textit{timer} based on \textit{batch}
  \If{\text{batchEnough}($i_{next}$, \textit{batch})} 
    \Comment{Check the size of \textit{batch}}
    \State Inference and reset \textit{batch} \Comment{Process the requests in \textit{batch}}
    \State \textit{timer}.clear()
  \ElsIf{\textit{timer}.isTimeout() \textbf{and} $|batch| > 0$}
    \State Inference and reset \textit{batch}
    \State \textit{timer}.clear()
  \EndIf
  \State \textit{req} $\gets$ \textit{queue}.pop() 
  \If{\textit{req} == $null$} 
    \If{$|batch| > 0$} 
      \State Inference and reset \textit{batch}
      \State \textit{timer}.clear()
    \Else
      \State \textbf{continue} 
    \EndIf
  \EndIf
  \If{\textit{req}.length < $i_{next}$}
    \State \textit{batch}.add(\textit{req})
    \If{$|batch| == 1$} \State \textit{timer}.start() \EndIf
  \Else
    \State Inference \textit{req}
  \EndIf
\EndLoop
\end{algorithmic}
\end{algorithm}
Eq.~\eqref{eq:prefill_total} explains why, for small batch sizes or small models, increasing $TP$ can increase $T_{p}$ when $\tau(TP)$ dominates the compute acceleration. \textcolor{black}{When $ISL$ is small, $\tau(TP)$ dominates $T_{p}(ISL,TP)$, and tensor parallelism may have a counterproductive effect. As $ISL$ increases, $C(ISL)$ grows faster than $\tau(TP)$ and thus comes to dominate $T_{p}(ISL,TP)$; in this regime increasing $S(TP)$ reduces $T_{p}(ISL,TP)$ and acceleration begins to manifest. However, on GPUs with lower inter-GPU communication bandwidth, $\tau(TP)$ constitutes a larger fraction of the total latency, the communication-dominated region expands, and} \textcolor{black}{tensor parallelism on a P-instance becomes less favorable. Conversely, on accelerators with high inter-GPU communication bandwidth, $\tau$ accounts for a smaller proportion, but the acceleration effect is still basically sublinear. To illustrate this effect empirically, we validated the behavior on small-model prefill experiments with varying $ISL$: when using Nvidia H100 GPUs to serve a QWen-1.5B model\cite{qwen}, for ${ISL}$ < 400 the configuration with $TP$ = 4 yields larger TTFT than the configuration with no tensor parallelism. Although the acceleration from tensor parallelism becomes more apparent as $ISL$ increases, the observed speedup (approximately 2.38) still falls well short of a linear $TP$ factor. So, for P-instance, 
adding new instances is far more efficient than increasing $TP$. Therefore, it is  recommended to set $TP$ only as large as required to support the model’s maximum input length.}


Under a high-concurrency mixed-length workload, a system configuration obtained from the average input and output lengths may become mismatched as lengths vary. To address this, short requests in the pending workload can be batched up to the input length corresponding to the system configuration so that the system adapts to the current load. To avoid degrading SLO attainment due to such batching, we should first model and analyze request queuing and timeout selection. To reason about queuing delay and timeout selection formally, let incoming requests follow an average arrival rate $\lambda$, and denote the mean prefill service time by $\mathbb{E}[S]$ (where $\mathbb{E}[S]=\mathbb{E}[T_{p}]$ given the chosen $TP$ and batching strategy). Then the system utilization is
\begin{equation}
\rho = \lambda \mathbb{E}[S].
\end{equation}
By Little's law the average number of requests in system $L$ and the mean sojourn time $W$ satisfy $L=\lambda W$.
For non-exponential arrival/service statistics, the (Kingman) heavy-traffic approximation gives the queueing waiting time $W_q$ same as Eq.~\eqref{eq:kingman}, where $c_a^2$ and $c_s^2$ are the squared coefficients of variation of inter-arrival and service times respectively. Eq.~\eqref{eq:kingman} and Eq.~\eqref{eq:prefill_total} allow us to (i) evaluate how a choice of $TP$ affects $\mathbb{E}[S]$ and hence $\rho$, and (ii) set the maximum waiting timeout $T_{timeout}$ to control the contribution of $W_q$ to TTFT. For example, a conservative choice is to enforce for a safety factor $\alpha\in(0,1]$ to bound expected extra waiting:
\begin{equation}
T_{{timeout}}\ge \alpha\cdot W_q.
\end{equation}

\begin{figure}
	\centering
	\includegraphics[width=\linewidth]{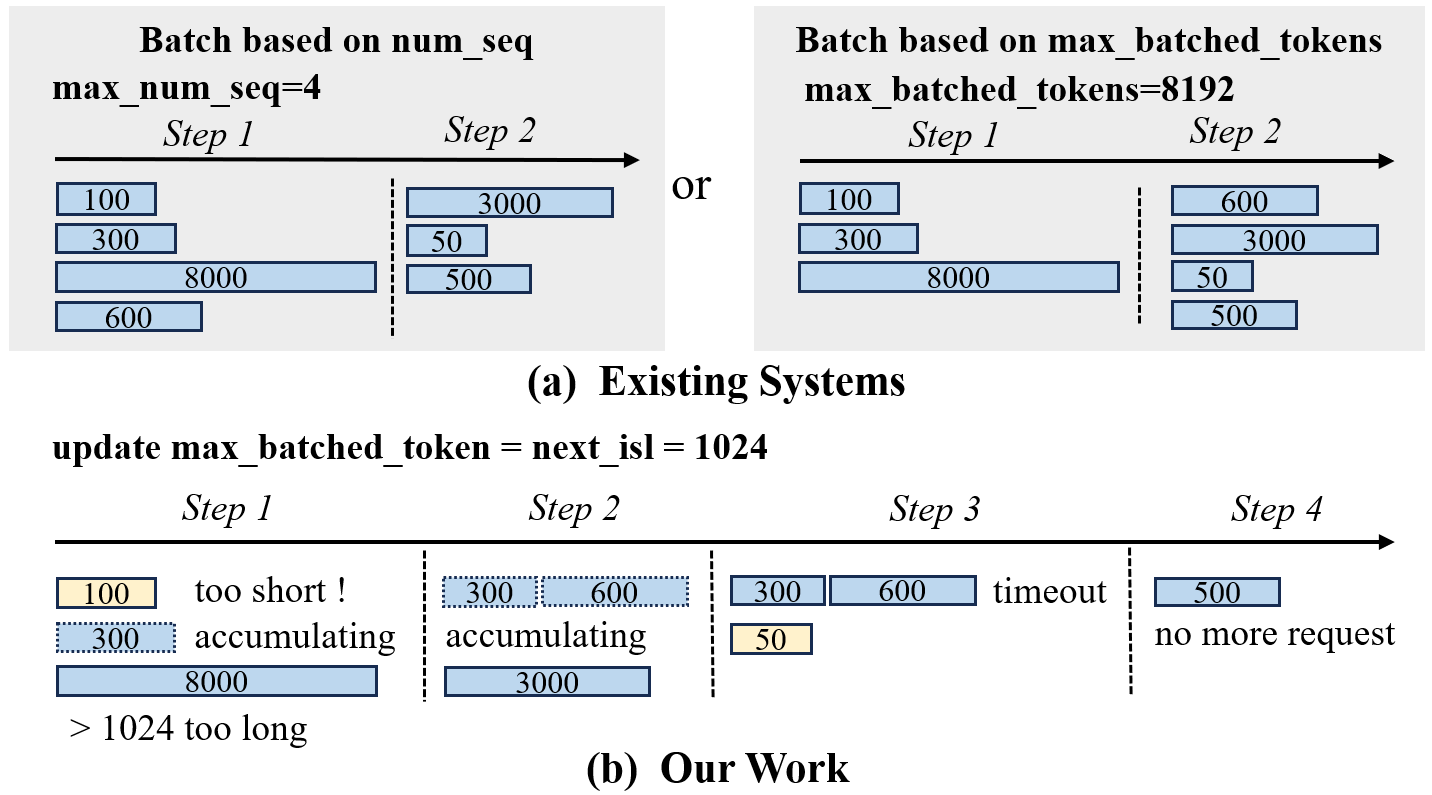}
	\caption{Comparison of prefill request scheduling solutions. All blocks represent the prefill inference process for sequences, and where the blue block represents the prefill inference for that sequence performed in a P-instance, and the yellow block represents the prefill inference performed in a D-instance. The solid block represents processing, and the dashed block represents unprocessed at the current step and still pending processing.}
	\label{fig:req_schedule_prefill}
\end{figure}
Because LLM request lengths vary widely, the conventional first-come-first-served batching policy, taking the first $n$ queued requests and forming a batch, can be suboptimal (as shown in Figure~\ref{fig:req_schedule_prefill}(a)). Such a policy may batch long and short requests together, causing long requests to push the GPU into a compute bottleneck and mismatch of system instances resource configuration, resulting in uneven resource utilization. This can be mitigated by a length-aware internal scheduling policy. When a P-instance dequeues requests from the prefill queue, it inspects each request's input length and applies the following logic.

If a request's length exceeds a predefined threshold, the P-instance dispatches it immediately for inference (as shown in step 1 and step 2 in the flow of Figure~\ref{fig:req_schedule_prefill}(b)). In DOPD, we set this threshold to the ARIMA-based predictor's one-step forecast of the next average input sequence length (denoted $next\_isl$ in Figure~\ref{fig:req_schedule_prefill}(b)), so that naturally long requests are not delayed by batching short ones. If a request is short, the instance accumulates short requests until either an accumulated-length constraint or a waiting-time constraint is met (as shown in step 1 and step 2 in Figure~\ref{fig:req_schedule_prefill}(b)). The batching-selection can be formalized as a (0\text{-}1) knapsack problem: given a set of candidate short requests indexed by $i$ with input lengths $w_i$ and benefit values $v_i$ (e.g., estimated throughput improvement or reduced per-request overhead), choose binary indicators $x_i\in\{0,1\}$ to
\begin{align}
\label{eq:knapsack}
\max_{\{x_i\}} &\quad \sum_i v_i x_i \\
\text{s.t.} &\quad \sum_i w_i x_i \le W_{\text{batch}}, \nonumber\\
&\quad x_i\in\{0,1\}, \nonumber
\end{align}
where $W_{\text{batch}}$ is the accumulated-length threshold (an operational counterpart in Algorithm~\ref{alg:prefill_request_scheduler} $i_{next}$). In online settings we recommend a greedy heuristic that orders candidates by $v_i/w_i$ (value per token) to maintain low scheduling overhead.

To avoid excessive waiting for short requests, a maximum waiting time is enforced. Once accumulated requests have waited longer than this timeout, the batch is executed immediately (as shown in step 3 in Figure~\ref{fig:req_schedule_prefill}(b)). Combining the above models—queueing delay estimates (\eqref{eq:kingman}), and knapsack-style batching selection (Eq.~\eqref{eq:knapsack})—yields a principled length-aware requests scheduler that (i) sets $T_{timeout}$ as a function of $W_q$ to bound TTFT contribution from queuing, and (ii) forms batches to maximize utilization while limiting per-request added waiting. Algorithm~\ref{alg:prefill_request_scheduler} implements this scheduling process. It can be observed that the scheduling routine incurs very low overhead. Its dominant cost lies in queue maintenance, so the time complexity per dequeue operation is $\mathcal{O}\big(N\big)$, where $N$ denotes the number of requests in the queue.

\subsection{Modeling of  D-instance}
The decoding stage is primarily constrained by memory rather than compute. Therefore, D-instances typically process multiple requests concurrently using batching to improve device utilization and system throughput. For LLM inference services, the D-instance batch size directly determines aggregate throughput: larger batch sizes generally yield higher throughput. The maximum feasible batch size for a D-instance is mainly determined by the GPU memory available for storing the KV-cache. Consequently, the distributed deployment parallelism for D-instances must be chosen carefully.

A prevalent strategy is to deploy D-instances using tensor parallelism because with $TP$ way tensor parallelism each GPU holds only $\frac{1}{TP}$ of the model weights, substantially reducing per-GPU memory footprint. This configuration both frees GPU memory for larger KV-cache and distributes computation across GPUs. When batch sizes are large, the relative communication overhead of tensor parallelism is small, making tensor parallelism deployment particularly suitable for D-instances. In general, increasing $TP$ allows a D-instance to support a larger maximum batch size and thus improves the system's ability to handle high-concurrency workloads. However, setting $TP$ excessively large may lead to resource waste. Guidelines for selecting an appropriate $TP$ for D-instances are provided in subsection~\ref{subsec:optimal_pd}.
\begin{figure}[t]
\centering
\includegraphics[width=\linewidth]{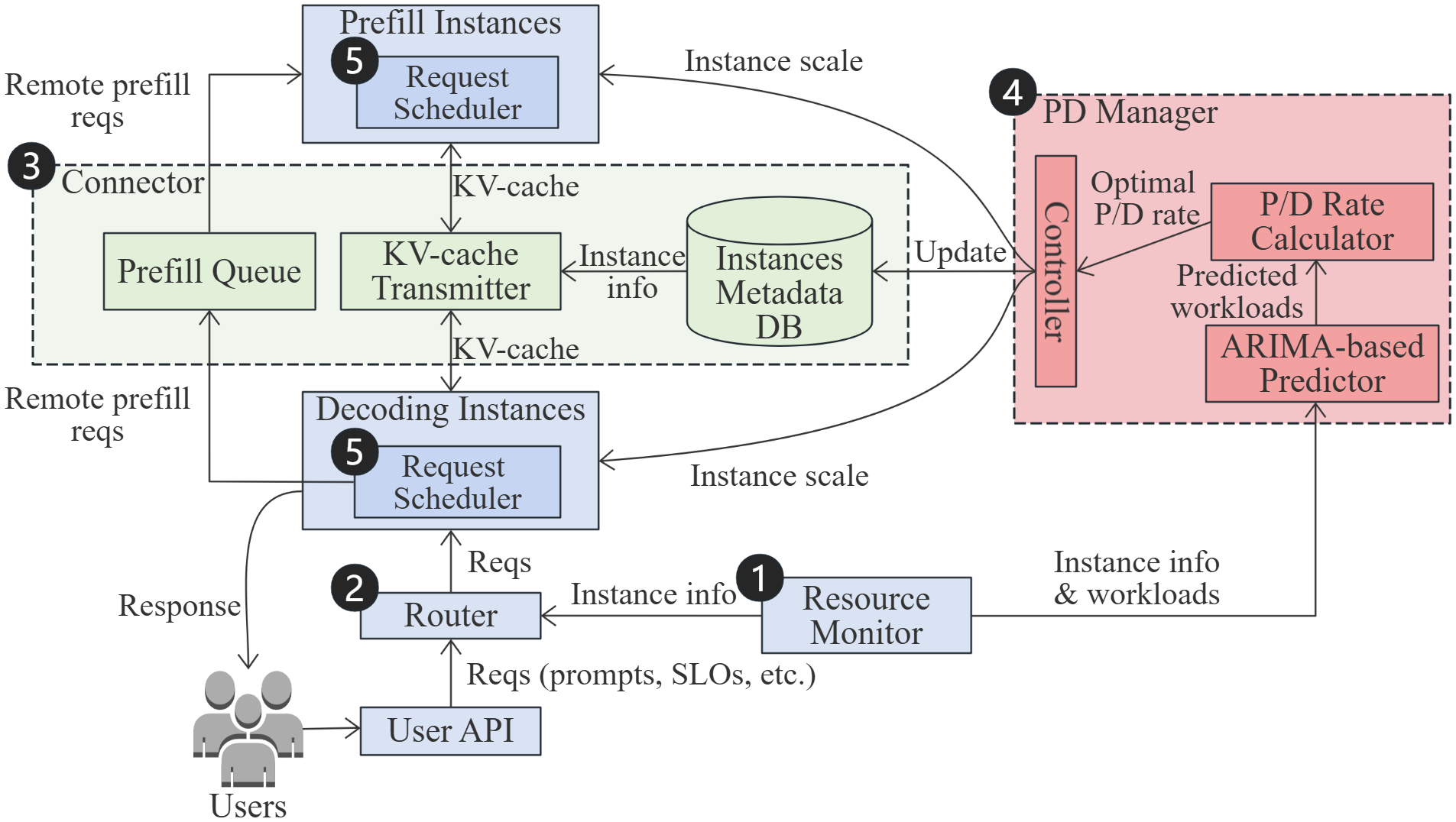}
\caption{DOPD system architecture.}
\label{fig:architecture}
\end{figure}
\section{SYSTEM DESIGN AND IMPLEMENTATION}
\label{sec:design}
We propose a method to accurately compute an appropriate P/D ratio based on current load and to dynamically reconfigure the P-instances and D-instances accordingly, so as to improve throughput with minimal GPU resources and meet the SLO attainment, thereby achieving higher goodput in PD-Disaggregation LLM inference systems. Building on these objectives, we extend the Dynamo framework to realize a dynamically optimal PD-Disaggregation inference system. The system comprises five principal components: (1) Resource Monitor, (2) Router, (3) Connector, (4) PD Manager, and (5) Request Scheduler.

\subsection{Resource Monitor}
The Resource Monitor (denoted by \blackcircnum{1} in Figure~\ref{fig:architecture}) is a core component responsible for collecting cluster- and instance-level performance metrics and load indicators to construct profiles for short-term load prediction and scheduling decisions. Both the Request Scheduler and the PD Manager rely on the Resource Monitor for up-to-date telemetry. Monitored signals include the number of P-instances, available GPU memory on D-instances, KV-cache utilization on D-instances, the size of the prefill queue, and so on. The Resource Monitor also records recent averages of TTFT, TPOT, ISL, and OSL with Prometheus\cite{prometheus} to meet the SLO attainment and to provide inputs used by the PD Manager and the Request Scheduler.

\subsection{Router}
The Router (denoted by \blackcircnum{2}) aggregates KV-cache information across all D-instances and routes incoming user requests to the most suitable instance. The Router is integrated with prefix-caching logic. For each request it computes the expected KV-cache hits on each D-instance, combines that information with per-instance load reported by the Resource Monitor, and preferentially forwards the request to the instance with the highest expected cache-hit count and the lowest inferred load.

\subsection{Connector}
The Connector (denoted by \blackcircnum{3}) provides inter-instance connectivity, maintains the queue of requests awaiting remote prefill, and implements end-to-end KV-cache transfers between P-instances and D-instances. The Connector contains three primary components: (1) an instances metadata DB, (2) a prefill queue, and (3) a KV-cache transmitter.
The instances metadata DB is implemented based on \text{etcd} service that registers the IDs of all P-instances and D-instances together with the addresses of their KV-cache blocks whenever a new instance joins the cluster.
The prefill queue is implemented using a \text{NATS} service acting as a queue that holds requests awaiting remote prefill (prefill inference in P-instance).
The KV-cache transmitter leverages NVIDIA's open-source NIXL\cite{nixl} communication library to perform high-performance GPU-to-GPU transfers of KV-cache blocks, thereby migrating KV-cache transfer latency as a dominant bottleneck in PD-Disaggregation.

\begin{algorithm}[t]
{\color{black}
\caption{Dynamic P/D Ratio Adjustment}\label{alg:pd_adjust}
\begin{algorithmic}[1]
\State \textbf{Input:}
\State $CC_{max}$: The D-instance's concurrency cap
\State \text{predictor}: ARIMA predictor
\State \text{interpolator}: Interpolator for TTFT, TPOT
\State $T$: Adjust interval
\State $t_{last}, t_{cur}$: Last record and current timestamps
\State $i_{last}, i_{next}$: Last and predicted input size
\State $o_{last}, o_{next}$: Last and predicted output size
\State $r_{last}, r_{next}$: Last and predicted request count
\State $TP_p$: Best tensor parallel size of P-instance
\State $TP_d$: Best tensor parallel size of D-instance
\State $ttft_{next}, tpot_{next}$: Interpolated performance metrics
\State $R_{opt}$: Optimal P/D (prefill/decoding) ratio
\State $n_p, n_d$: Next number of P-instance and D-instance
\State \text{adjustPD}: Instance adjustment action
\State \textbf{Initialize} \text{predictor}, \text{interpolator}, $t_{last}$
\Loop
  \State update $r_{wait}, load_{kv}, t_{cur}$
  \If{\text{checkOverload}($r_{wait}, load_{kv}$)}
    \State $n_p \gets \lceil r_{wait}/SIZE\_QUEUE \rceil$
    \State $n_d \gets \lceil load_{kv} / KV\_CAP \lceil$
    \State \text{adjustPD}($n_d, TP_p, n_p, TP_d$)
  \EndIf
  \If{$t_{cur} - t_{last} > T$}
    \State update $i_{last}, o_{last}, r_{last}$
    \State $t_{last} \gets t_{cur}$
    \State \text{predictor.add}($i_{last}, o_{last}, r_{last}$)
    \Statex ARIMA one-step forecast (where $p, q, \mu, \varphi, \theta, \hat\varepsilon$ denote ARIMA parameters and estimated residuals):
    \[ \hat X_{t+1\mid t} = \mu + \sum_{j=1}^p \varphi_j X_{t+1-j}
                         + \sum_{k=1}^q \theta_k \hat\varepsilon_{t+1-k} \]
    \State $i_{next}=\hat I_{t+1\mid t}, o_{next}=\hat O_{t+1\mid t}, r_{next}=\hat R_{t+1\mid t}$
    \State $ttft_{next}$ $\gets$ \text{interpolator}($i_{next}$)
    \State $tpot_{next}$ $\gets$ \text{interpolator}($r_{next}, \frac{i_{next}+o_{next}}{2}$)
    \State $R_{opt} \gets \dfrac{r_{next}}{(o_{next} \times tpot_{next}) / ttft_{next}}$
    \State $n_d \gets \lceil r_{next} / CC_{max} \rceil$
    \State $n_p \gets R_{opt} \times n_d$
    \State \text{adjustPD}($n_p, TP_p, n_d, TP_d$)
  \EndIf
\EndLoop
\end{algorithmic}
}
\end{algorithm}
\subsection{PD Manager}
\label{subsec:pd_manager}
The PD Manager (denoted by \blackcircnum{4}) is a component that can calculate the optimal P/D ratio based on historical load and adjust the instances in the system to this ratio. 
To improve the fidelity of the $P/D$ ratio computation, it is essential to perform pre-deployment profiling of the target model and GPU. Specifically, we first profile a single P-instance across a range of input sequence lengths to obtain accurate TTFT measurements as a function of input length. Similarly, we profile a single D-instance across combinations of sequence lengths and batch sizes to obtain precise TPOT measurements as a function of sequence length and batch size. 
These profiles guide parallel-configuration choices during scaling and provide expected TTFT/TPOT values for given sequence lengths and concurrency. "Initialize interpolator" in Algorithm~\ref{alg:pd_adjust} performs this process. \textcolor{black}{This profiling overhead is determined by the model size and the GPU specifications; generally, larger model dimensions or lower GPU compute power lead to increased profiling overhead. Taking the combination of the NVIDIA H100 GPU and the LLaMa-3.3-70B-FP8 model as an example, the profiling overhead for TTFT is approximately 90 minutes, while the profiling overhead for TPOT is approximately 150 minutes.} 

At runtime, the Resource Monitor maintains a recent distribution of request lengths. To avoid the overhead of expensive predictions at scale, we employ a lightweight time-series model (ARIMA) to predict the future average input and output lengths. We compare the predicted values with the recently observed averages from the Resource Monitor to derive correction factors \textcolor{black}{(defined as the ratio of the predicted value to the ground-truth value, used to calibrate subsequent predictions and update ARIMA parameters; this enables the ARIMA predictor to adapt more rapidly to new workloads during periods of high volatility)}. To reduce forecast error, we combine the derived correction factors with the 
predicted average sequence lengths and the profiled performance parameters to compute the projected optimal P/D ratio. \textcolor{black}{Practical testing shows that ARIMA's average latency is under 0.8s, a negligible overhead compared to the hundred-second reconfiguration intervals. Under real-world workloads, ARIMA maintains over 87\% accuracy (defined by errors within 10 requests for future concurrency and 50 tokens for $ISL$ and $OSL$) failing only during rare bursty conditions ($<5\%$). To handle such unpredictable spikes, the system adaptively monitors prefill queue depth ($r_{queue}$) and KV-cache occupancy ($load_{kv}$). Upon detecting imminent overload, the system bypasses the scheduled interval and immediately scales P-instances to $r_{queue}/SIZE\_QUEUE$ and D-instances to $load_{kv}/KV\_CAP$, where the $SIZE\_QUEUE$ denotes the size of prefill queue, and $KV\_CAP$ denotes the maximum adjustment threshold of KV-cache memory utilization, as detailed in Algorithm~\ref{alg:pd_adjust} (lines 18–23). To further mitigate resource wastage caused by occasional over-prediction, we pre-define a maximum adjustment threshold for the number of instances, constraining the scale of individual reconfiguration operations.}


Algorithm~\ref{alg:pd_adjust} summarizes the procedure for forecasting future load, deriving the optimal P/D ratio, and executing instance adjustments. DOPD first jointly considers the factors listed above to determine the optimal tensor parallel size $TP$ for P-instances and D-instances, and then performs instance adjustments according to the computed optimal P/D ratio. This design enables accurate forecasting of a future, workload-appropriate P/D ratio from historical telemetry and supports non-disruptive (zero-downtime) elastic resizing of P-instances and D-instances to promptly accommodate time-varying user request loads. 

\begin{figure}
\centering
\includegraphics[width=\linewidth]{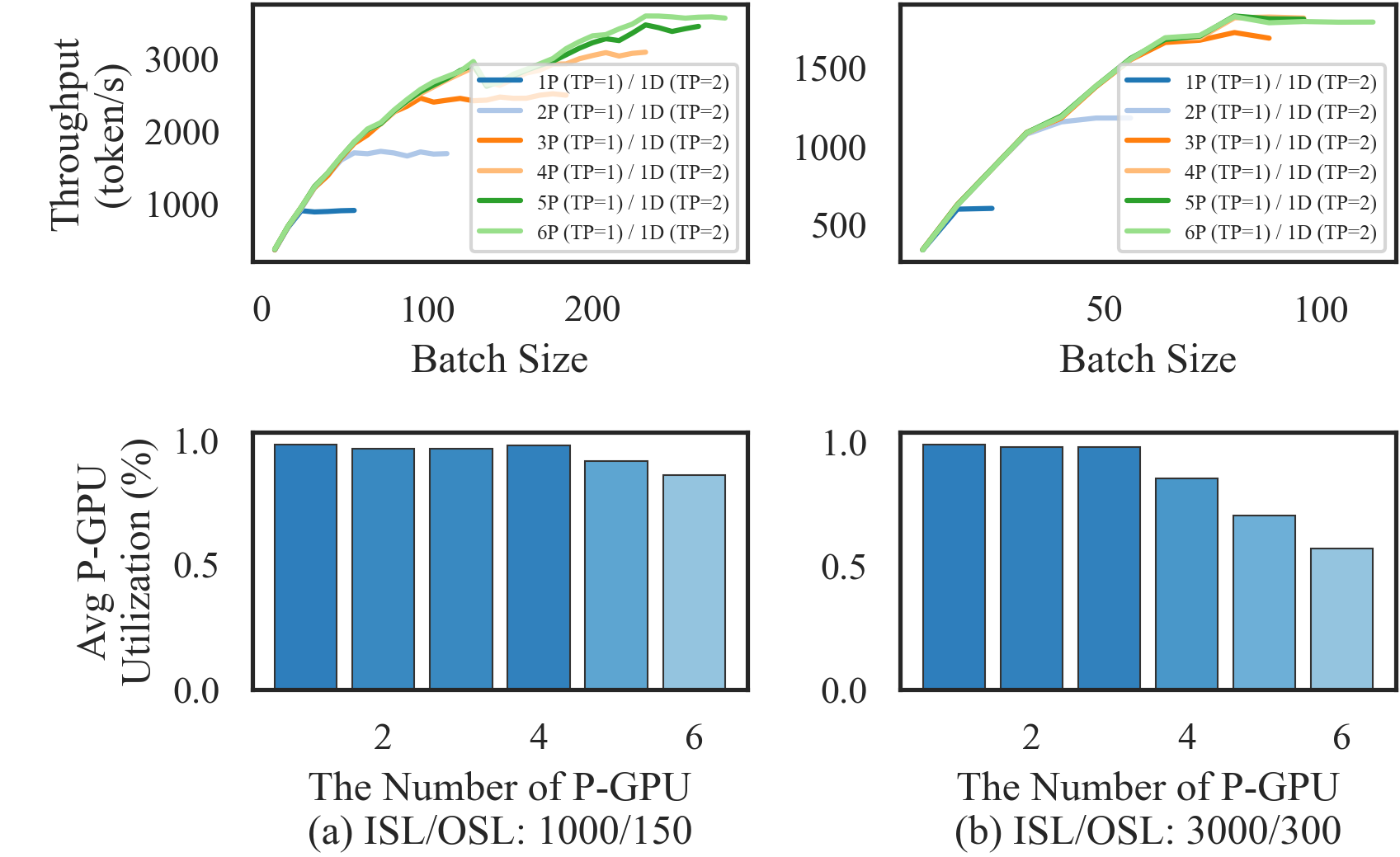}
\caption{Optimal P/D ratio validation experiment results for varying numbers of P-instances and a single D-instance with $TP = 2$ under (a) fixed input/output sequence length of $1000/150$ (optimal P/D ratio is approximately 4.58). (b) fixed input/output sequence length of $3000/300$ (optimal P/D ratio is approximately 3.18).}
\label{fig:optimal_pd_verify}
\end{figure}
\subsection{Request Scheduler}
The Request Scheduler (denoted by \blackcircnum{5}) is built into P-instances and D-instances to arrange requests reasonably, allowing the PD-Disaggregation system to better cope with mixed requests. To implement this, we group and batch requests by their $ISL$. Under high concurrency, short requests are batched according to their predicted future lengths to leverage kernel efficiency, reducing the mismatch between short-length load and the provisioned P/D ratio. Long requests, whose computational cost is inherently high, are not merged and are dispatched to P-instances for immediate execution.

For ultra-short requests we employ PD-aggregation: merging their prefill with other requests' decoding on D-instances avoids unnecessary KV-cache transfers and relieves prefill queue pressure (illustrated by the yellow blocks in step 1 and step 3 of the flow in Figure~\ref{fig:req_schedule_prefill}(b)). Using this request-scheduling scheme, our experiments on realistic workloads demonstrate substantial improvements in maximum concurrency, throughput, and TTFT without changing the global P/D ratio. Detailed results appear in Section~\ref{sec:eval}.

\section{PERFORMANCE EVALUATION}
\label{sec:eval}
This section carries out experimental evaluation of the proposed DOPD system. It discusses configuration and analyses of our experiments and results along with scalability of DOPD and the directions of future research.
\subsection{Evaluation Setup}

\subsubsection{Testbed}
We conduct experiments on a local cluster composed of $8$ NVIDIA H100-SXM-80GB\cite{H100} GPUs.

\subsubsection{LLM models}
For inference-performance evaluation we select three representative models used in prior works and industries test suites: \text{OPT-30B}\cite{opt-30b},  LLaMa-3.3-70B-FP8\cite{DeepSeek-R1-Distill-Llama-70B-FP8-dynamic} and QWen2.5-72B-FP8\cite{QWen2.5-72B-FP8}.

\subsubsection{Datasets}
\textcolor{black}{We employ three production workload traces collected from Microsoft (BurstGPT\cite{burstgpt}, Azure code Traces\cite{AzurePublicDataset} and Azure conversation\cite{AzurePublicDataset}), and we use the ShareGPT\cite{sharegpt} dataset for request contents. These traces provide realistic workload characteristics.}

\subsubsection{Resource monitoring}
To collect and aggregate telemetry we leverage NVIDIA's NVML\cite{nvml} for low-level GPU metrics and Prometheus\cite{prometheus} for metrics aggregation and retention.


\begin{figure}
    \centering
    \includegraphics[width=\linewidth]{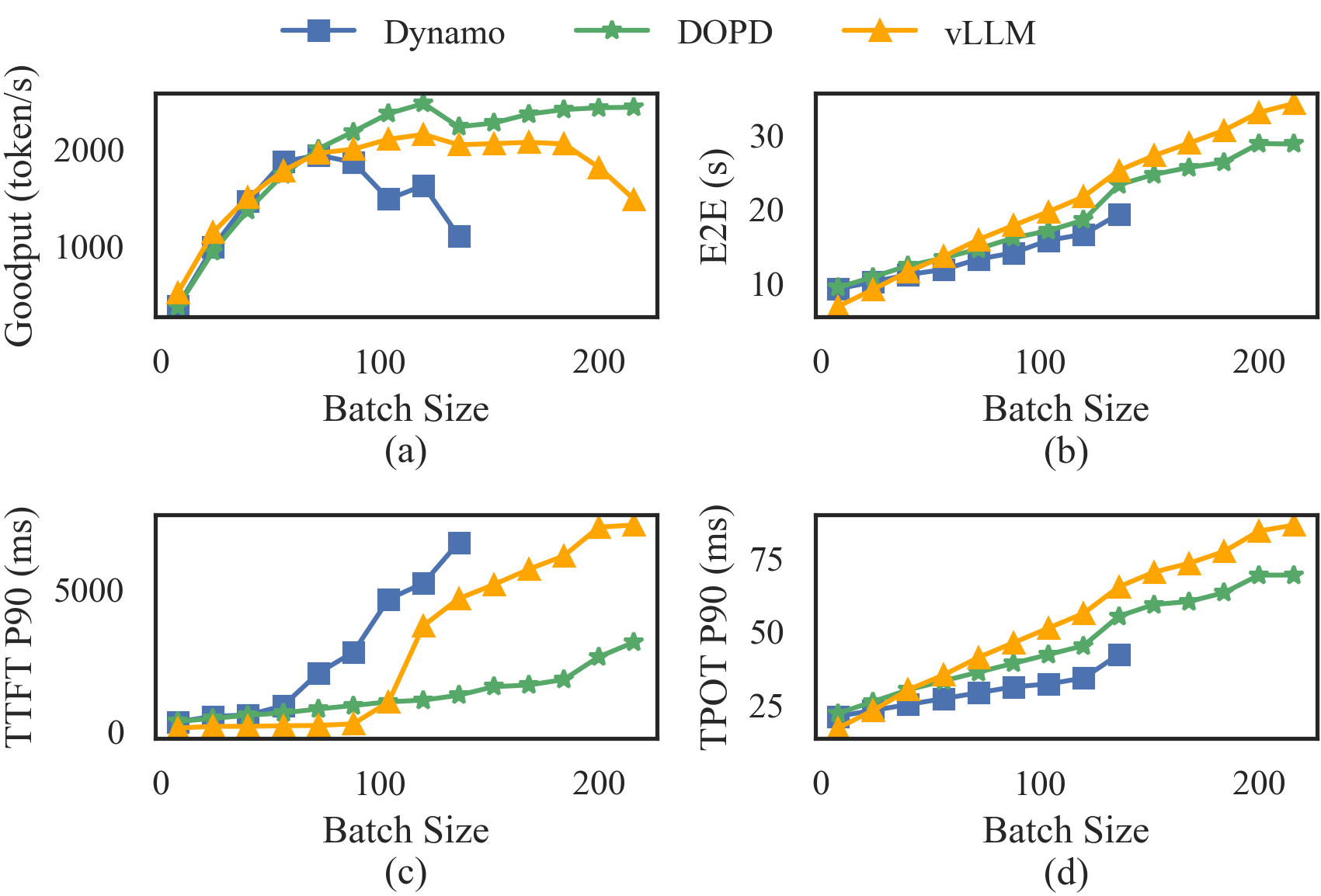}
    \caption{The performance comparison in terms of (a) goodput, (b) end-to-end latency, (c) TTFT, and (d) TPOT of a static P/D ratio on real Azure workload experiments with LLaMa-3.3-70B-FP8 model, compared with aggregated methods.}
    \label{fig:static_comp_70b}
\end{figure}
\subsubsection{Baselines}
We divide the experiment into static experiments and dynamic experiments.

\paragraph{Static experiments} Under a fixed P/D configuration we compare the following systems:
\begin{itemize}
\item \textbf{DistServe}\cite{DistServe}: a representative PD-Disaggregation framework with strong performance characteristics.
\item \textbf{Dynamo}\cite{ai-dynamo}: the most feature-complete open-source PD-Disaggregation framework.
\item \textbf{vLLM}\cite{vllm}: a state-of-the-art open-source LLM inference system with broad adoption.
\end{itemize}

\paragraph{Dynamic experiments.} \textcolor{black}{Because mature dynamic PD-Disaggregation approaches are not open-source and widely available apart from Dynamo, we compare DOPD against two Dynamo scheduling strategies}:
\begin{itemize}
\item \textbf{DYN-LOAD}: a load-based scheduler that scales based on runtime utilization thresholds.
\item \textbf{DYN-SLA}: an SLA-aware scheduler that derives scaling decisions from real-time inference monitor metrics.
\end{itemize}

\subsection{Evaluation Results}
\subsubsection{Validation of the Optimal P/D Ratio}
\textcolor{black}{To validate the correctness of the computed Optimal P/D Ratio}, we first deploy Dynamo with the \text{LLaMa-3.3-70B-FP8} model, varying the number of P-instances (each on $1$ H100 GPU) while fixing the number of D-instances to one (with $TP = 2$ across $2$ H100 GPUs). Prior to service launch, our computation yields an optimal P/D ratio of approximately $4.58$, indicating that roughly $4.58$ P-instances suffice to balance production and consumption.

During each experiment we issue inference requests with fixed input length $ISL=1000$ and output length $OSL=150$ at varying concurrency levels (concurrency is defined as the number of in-flight requests in system). Figure~\ref{fig:optimal_pd_verify}(a) illustrates throughput under the fixed $1000/150$ workload. It shows a near-linear increase in maximum throughput as the number of P-instances increases from $1$ to $4$. Beyond $4$, throughput growth from $4$ to $5$ slows markedly, and from $5$ to $6$ is negligible. Concurrently, the average GPU utilization of P-instances (as shown in the 
lower figure in Figure~\ref{fig:optimal_pd_verify}(a)) indicates idle GPU time only when the number of P-instances exceeds $4$. This proves that our pre-computed optimal number of P-instances is correct (between 4 and 5).

We further compute that for input/output sequence length of $3000/300$, the corresponding optimal number of P-instances is $3.18$. Repeating the experiment under the $3000/300$ workload produces analogous results (as shown in Figure~\ref{fig:optimal_pd_verify}(b)), further confirming the accuracy of our calculation method.

\begin{figure}
    \centering
    \includegraphics[width=\linewidth]{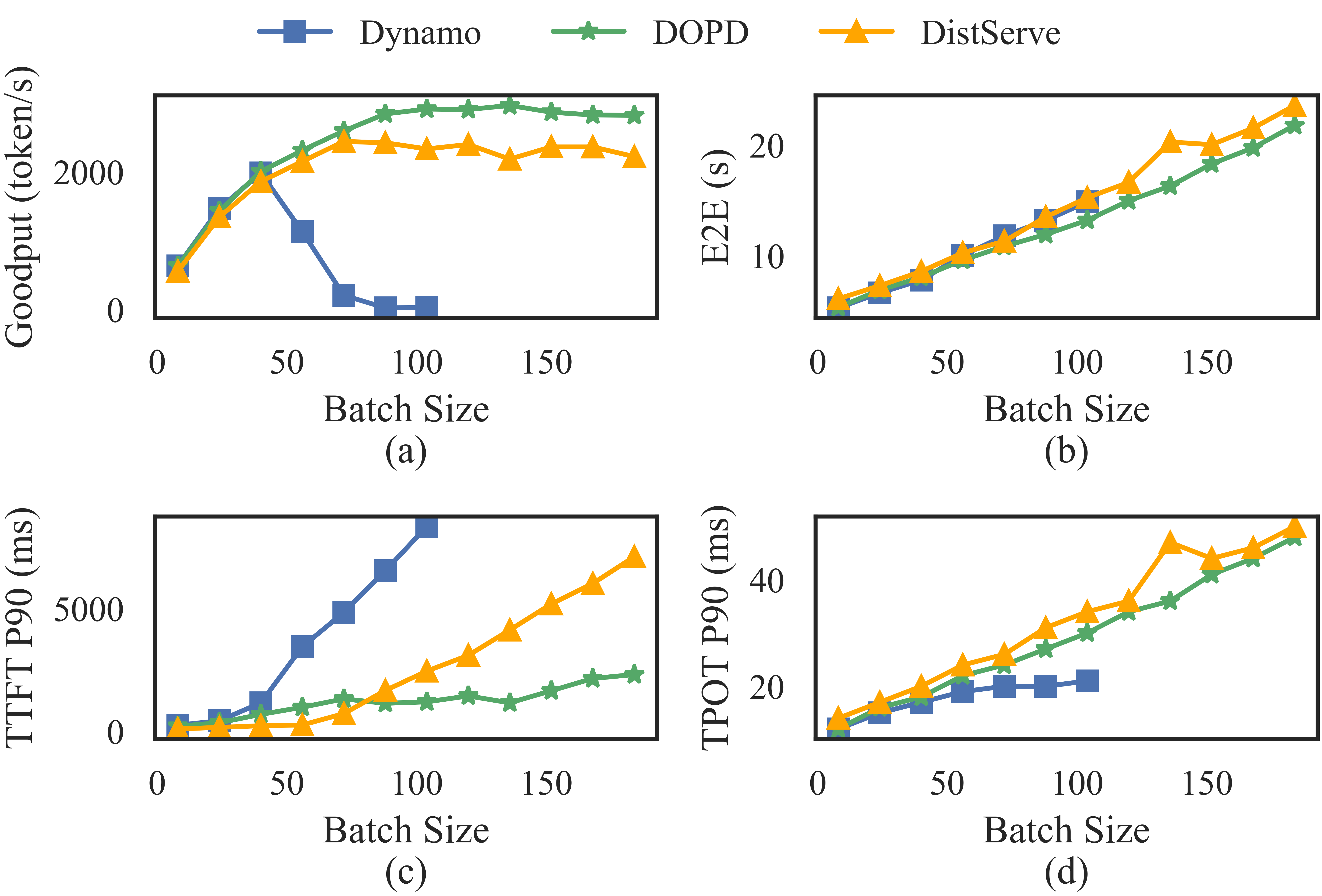}
    \caption{The performance comparison in terms of (a) goodput, (b) end-to-end latency, (c) TTFT, and (d) TPOT of a static P/D ratio on real Azure workload experiments with OPT-30B model, compared with disaggregated methods.}
    \label{fig:static_comp_30b}
\end{figure}
\subsubsection{Comparison of Static P/D Inference-Performance}
\textcolor{black}{To compare the performance of DOPD with representative PD-aggregated and PD-disaggregated approaches under a static P/D configuration,} we extract the first $1024$ entries from \text{Microsoft Azure traces} and use \text{ShareGPT} as the source of request content, generate a request set with average input/output sequence length of approximately $1024/245$. We then show two sets of experiments.

First, we compare DOPD with a representative PD-aggregation system and a PD-Disaggregation baseline using the \text{LLaMa-3.3-70B-FP8} model. The aggregation baseline (\text{vLLM}) is deployed on $4$ H100 GPUs, whereas both disaggregation baselines (\text{Dynamo} and \text{DOPD}) are configured as \text{2P (TP=1) / 1D (TP=2)}. We generate inference workload from the \text{Microsoft Azure traces} and replay it to each deployment. Results are reported in Figure~\ref{fig:static_comp_70b}, which shows goodput, TTFT, TPOT, and end-to-end latency under the same workload. Compared to 
vLLM, DOPD reduces P90 TPOT from $0.097$s up to $0.079$s and increases goodput by up to $1.5\times$. Under reasonable SLO constraints (violation rate $<5\%$), DOPD also supports a substantially higher maximum concurrency, rising from $72$ to $112$.

\begin{figure}
\centering
\includegraphics[width=\linewidth]{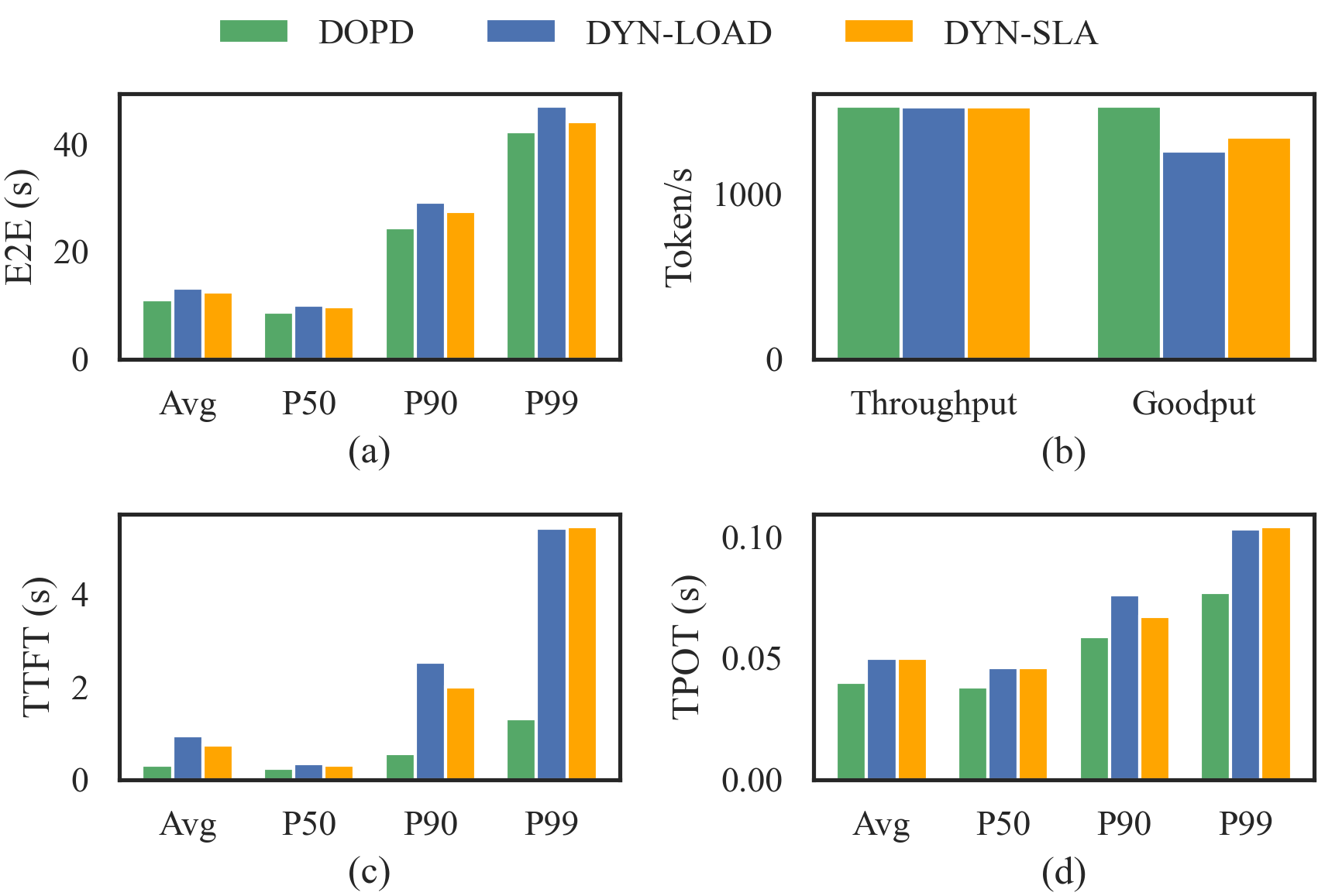}
\caption{The performance comparison in terms of (a) end-to-end latency, (b) throughput and goodput, (c) TTFT, and (d) TPOT of dynamic scheduling experiments under real-world BurstGPT workloads.}
\label{fig:dynamic_comp}
\end{figure}

\begin{figure}
\centering
\includegraphics[width=\linewidth]{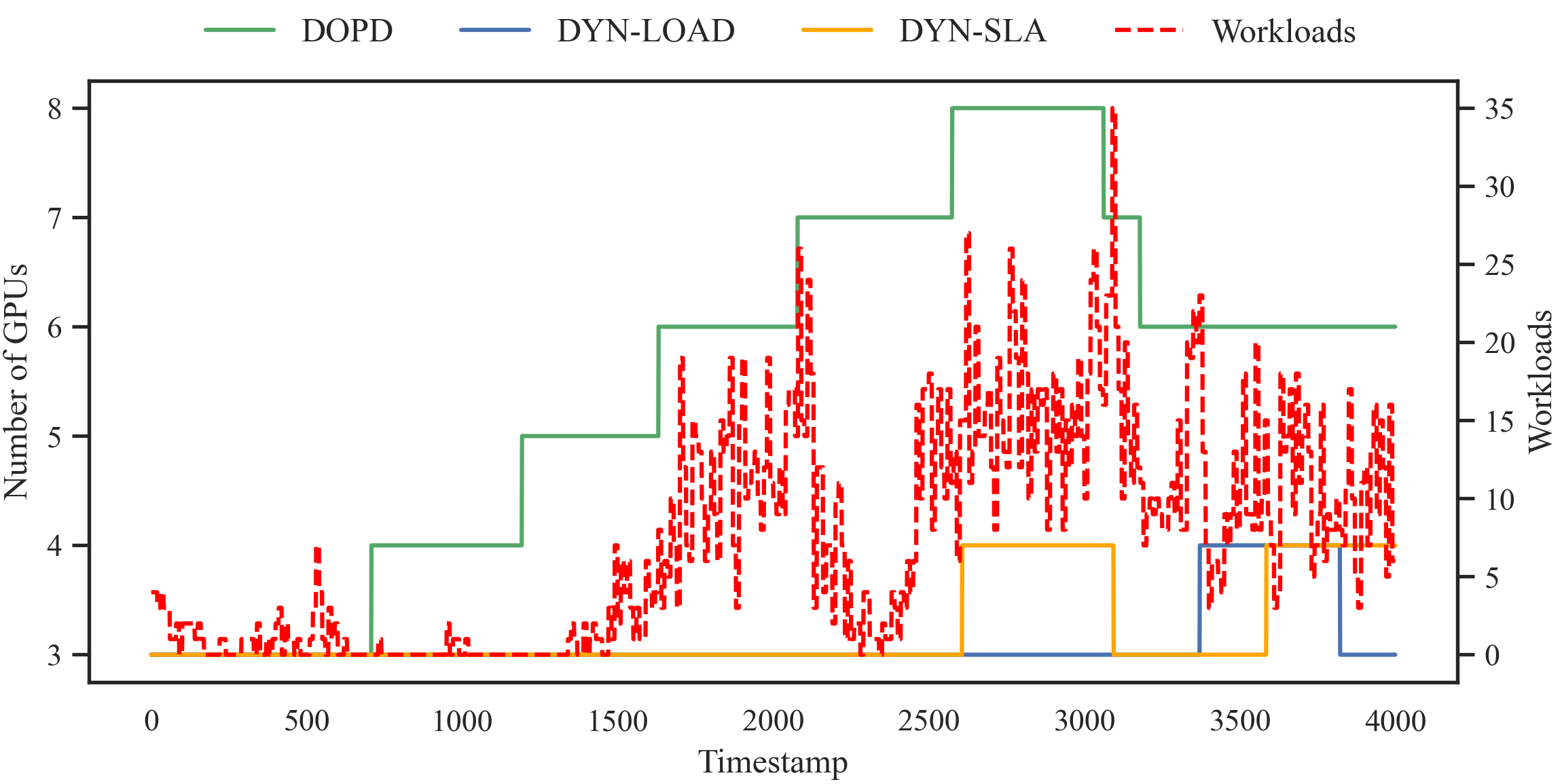}
\caption{GPU provisioning and workload variation over time. Solid lines (left axis) denote the GPU counts for DOPD, DYN-LOAD, and DYN-SLA, while the dashed line (right axis) represents the instantaneous request rate.}
\label{fig:dynamic_comp_num_gpu}
\end{figure}
Second, we evaluate three PD-Disaggregation systems using the \text{facebook/opt-30b} model. In these experiments each system  are configured as \text{1P (TP=1) / 1D (TP=4)}. Results appear in Figure~\ref{fig:static_comp_30b}. As concurrency increases, DOPD's benefits become pronounced: relative to a representative disaggregation system (\text{DistServe}), DOPD reduces P90 TTFT from $7.10$s to $2.31$s and improves goodput by up to 27\%. Under the SLO constraints (violation rate $<5\%$), the maximum supported concurrency increases from $88$ to $104$. 
Through DOPD's length-aware scheduling, it significantly outperforms Dynamo across all evaluation metrics.

\subsubsection{Dynamic P/D Instance-Scheduling Performance Comparison}
\textcolor{black}{To test the performance of DOPD and other dynamic instance-scheduling methods,} DOPD significantly outperforms Dynamo's built-in load-based and SLA-aware scheduling strategies under complex and highly variable inference workloads. We combine the workload of the BurstGPT dataset with the request content of the ShareGPT dataset to generate a simulated LLM inference workload. According to the timestamps in the BurstGPT trace, we replay requests to three service deployments. All three systems (DOPD, \text{DYN-LOAD}, \text{DYN-SLA}) are initially configured with one P-instance ($TP = 1$) and one D-instance ($TP = 2$) running the LLaMa-3.3-70B-FP8 model.
Figure~\ref{fig:dynamic_comp_num_gpu} shows the number of requests versus time (sampled every $0.1$s), while the other curves plot the total number of GPUs provisioned over time under each scheduling strategy. Although \text{DYN-SLA} and \text{DYN-LOAD} adjust resource allocation in response to load changes, their scaling amplitudes and timeliness are insufficient. In contrast, DOPD's GPU count closely tracks the RPS curve. It scales out rapidly during traffic spikes and scales in promptly when load subsides. Figure~\ref{fig:dynamic_comp} presents the results. From the four plots it is clear that DOPD's end-to-end (E2E) latency, TTFT, TPOT, and goodput all outperform the other baselines. As a result, DOPD achieves excellent SLO attainment (99.4\%), whereas \text{DYN-SLA} and \text{DYN-LOAD} incur worse SLO attainment ($87.3\%$ and $80.8\%$ respectively). These results demonstrate that DOPD can judiciously scale P-instances and D-instances in response to load fluctuations, substantially increasing the SLO attainment.

\begin{figure}
\centering
\includegraphics[width=\linewidth]{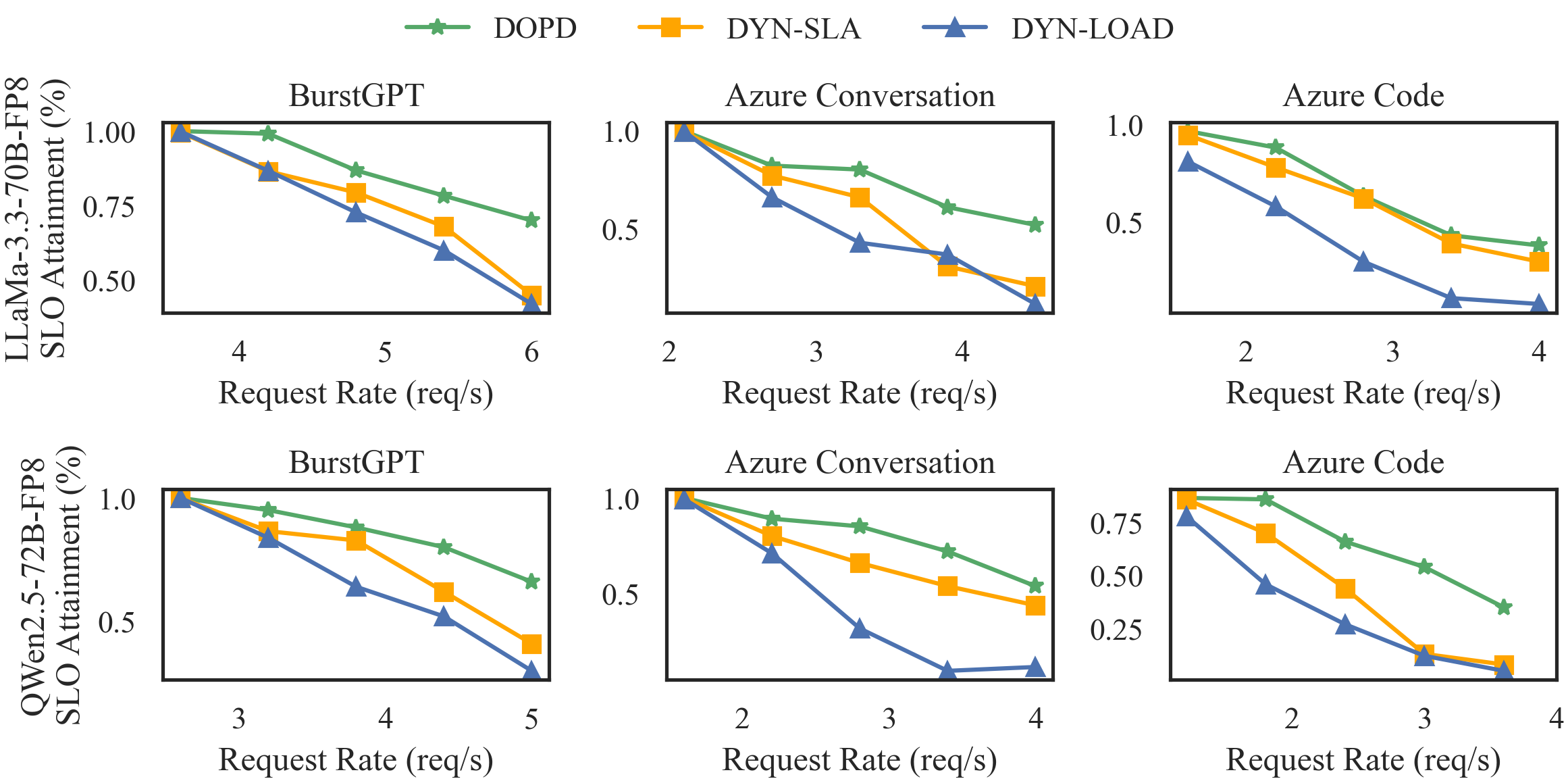}
\caption{\textcolor{black}{SLO attachment performance of different workload datasets under different requests rate and different scheduling policies for different models.}}
\label{fig:dynamic_rps}
\end{figure}
\textcolor{black}{To further demonstrate the versatility and robustness of our approach across diverse workloads and model architectures, we conducted extensive evaluations using traces derived from the BurstGPT, Azure Conversation, and Azure Code datasets. As illustrated in Figure~\ref{fig:dynamic_rps}, we evaluated the end-to-end performance under varying request rates for both LLaMa-3.3-70B-FP8 and QWen2.5-72B-FP8 models. For the LLaMa-3.3-70B-FP8 model, although all three scheduling strategies exhibit a decline in SLO attainment at high request rates due to limited hardware resources, DOPD consistently achieves the most graceful degradation, maintaining the highest SLO attainment across all datasets. In contrast, the DYN-LOAD baseline performs poorly, particularly under high concurrency, as its reactive adjustment strategy only triggers reconfiguration after the system has already reached an overloaded state. This latency prevents the system from proactively adapting to traffic surges, resulting in severe SLO violations. Similarly, for the QWen2.5-72B-FP8 model, DOPD maintains its superior performance with the slowest rate of SLO decline. Due to the lack of guidance for computing an optimal P/D ratio, DYN-SLA exhibits highly unstable performance and struggles to balance P-instances and D-instances, which in turn leads to a high rate of SLO violations. Notably, all methods face challenges when processing workloads with a high proportion of long-context requests. This performance bottleneck stems from both hardware quantity constraints and the inherent computational intensity of long-sequence prefill, highlighting a critical area for future research focused on optimizing tail latency in long-context inference scenarios.}

\section{\textcolor{black}{Discussion}}
\label{sec:discussion}
\textcolor{black}{\textbf{The cost of scheduling and scalability}. DOPD demonstrates high efficiency and scalability with minimal scheduling overhead. The PD Manager's execution, dominated by ARIMA forecasting, completes within 1s. As this process is asynchronous and occurs at infrequent intervals (hundreds of seconds), the resulting CPU overhead is negligible. Instance reconfiguration is equally efficient: graceful shutdown typically finishes within 5s, while scaling up a 70B model takes $\le 30$s for a P-instance (and $\approx 1.5\times$ longer for a D-instance due to CUDA Graph loading). Crucially, accurate workload prediction enables proactive scaling, effectively masking these initialization latencies. Furthermore, the Request Scheduler incurs only microsecond-level latency, and the instance-decoupled architecture inherently facilitates seamless scaling to large-scale clusters.}

\textcolor{black}{\textbf{Limitation}. Despite DOPD's robustness on realistic production traces, occasional ultra-long prompts can transiently degrade overall system performance. How to rapidly and efficiently process ultra-long context requests has remained a formidable challenge since the emergence of LLMs. 
Furthermore, in most current LLM inference scenarios, ultra-long context requests are rare and therefore difficult for the predictor to anticipate. Scheduling decisions that rely on average-length signals may perform poorly when extremely long contexts suddenly appear.
A potential and straightforward solution is to deploy an auxiliary specialized node dedicated to prioritizing ultra-long requests. 
Another limitation of our work is that the current approach does not yet support the dynamic selection of different $TP$ degrees for varying workloads, as this would introduce significantly more complex P/D ratio calculations and more intricate logic for instance reconfiguration. However, the system inherently supports the scaling or removal of instances with different $TP$ configurations. In the future, we can further refine our algorithm by modeling these more complex scenarios.} 

\textcolor{black}{\textbf{Future work and extension}.  Future work can explore targeted mitigations for ultra-long prompt interference and investigate integrations with sparse attention mechanisms or speculative-decoding architectures to enhance generality and tail robustness. Meanwhile, as user demand for Multimodal Large Language Models (MLLMs) increases, we also consider further extending our methodology by integrating it with emerging techniques (such as those described in \cite{EPD} and \cite{ModServe}) to optimize the inference performance of MLLMs. Nevertheless, DOPD provides a principled foundation for dynamic resource management in disaggregated LLM inference, striking a balance between scalability and operational practicality.}
\section{Conclusions}
\label{sec:conclusion}
In this paper, we proposed DOPD, a dynamic framework that combines PD-Disaggregation with dynamic adjustment of instances to improve goodput, reduce inference latency, achieve more stringent SLOs, and conserve GPU resources. We introduce a method for calculating the optimal P/D ratio and a complementary request scheduling strategy that enhances the adaptation of disaggregated systems to mixed-length workloads. The proposed P/D ratio calculation together with workload-aware scheduling enables DOPD to achieve excellent performance under complex, time-varying workloads.
Through extensive experiments on realistic production traces and representative LLMs, we demonstrate that DOPD achieves up to $1.5\times$ improvement in goodput while simultaneously reducing inference latency. DOPD also achieves a near-zero SLO violation rate under real-world workloads and reliably detects load fluctuations to trigger timely elastic scaling. Our results indicate that DOPD substantially improves the inference efficiency of PD-Disaggregation architectures for LLMs, thereby facilitating their deployment in industrial environments with demanding availability and performance requirements.

\section{Software Availability}
The codes of DOPD have been open-sourced to \url{https://github.com/liao4s/DOPD} for research usage.
\bibliographystyle{IEEEtran}
\bibliography{refs}

@IEEEtranBSTCTL{IEEEexample:BSTcontrol,
  CTLuse_forced_etal       = "yes",
  CTLmax_names_forced_etal = "3",  
  CTLnames_show_etal       = "3"   
}

@ARTICLE{BrownoutServe2025,
author={Hu, Jianmin and Xu, Minxian and Ye, Kejiang and Xu, Chengzhong},
journal={ IEEE Transactions on Computers },
title={{ BrownoutServe: SLO-Aware Inference Serving under Bursty Workloads for MoE-based LLMs }},
year={2026},
volume={},
number={01},
ISSN={1557-9956},
pages={1-14},
doi={10.1109/TC.2026.3655019},
url = {https://doi.ieeecomputersociety.org/10.1109/TC.2026.3655019},
publisher={IEEE Computer Society},
address={Los Alamitos, CA, USA},
month=jan}

@misc{qwen,
      title={Qwen Technical Report}, 
      author={Jinze Bai and Shuai Bai and Yunfei Chu and Zeyu Cui and Kai Dang and Xiaodong Deng and Yang Fan and Wenbin Ge and Yu Han and Fei Huang and Binyuan Hui and Luo Ji and Mei Li and Junyang Lin and Runji Lin and Dayiheng Liu and Gao Liu and Chengqiang Lu and Keming Lu and Jianxin Ma and Rui Men and Xingzhang Ren and Xuancheng Ren and Chuanqi Tan and Sinan Tan and Jianhong Tu and Peng Wang and Shijie Wang and Wei Wang and Shengguang Wu and Benfeng Xu and Jin Xu and An Yang and Hao Yang and Jian Yang and Shusheng Yang and Yang Yao and Bowen Yu and Hongyi Yuan and Zheng Yuan and Jianwei Zhang and Xingxuan Zhang and Yichang Zhang and Zhenru Zhang and Chang Zhou and Jingren Zhou and Xiaohuan Zhou and Tianhang Zhu},
      year={2023},
      eprint={2309.16609},
      archivePrefix={arXiv},
      primaryClass={cs.CL},
      url={https://arxiv.org/abs/2309.16609}, 
}

@online{Qwen2.5-72B-FP8,
  author  = {RedHatAI},
  title   = {RedHatAI/Qwen2.5-72B-Instruct-FP8-dynamic},
  year    = {2024},
  url     = {https://huggingface.co/RedHatAI/Qwen2.5-72B-Instruct-FP8-dynamic},
  urldate = {2024-12-02},
  note = {Accessed: 2026-01-05}
}

@inproceedings{
EPD,
title={Efficiently Serving Large Multimodal Models Using {EPD} Disaggregation},
author={Gursimran Singh and Xinglu Wang and Yifan Hu and Timothy Tin Long Yu and Linzi Xing and Wei Jiang and Zhefeng Wang and Bai Xiaolong and Yi Li and Ying Xiong and Yong Zhang and Zhenan Fan},
booktitle={Forty-second International Conference on Machine Learning},
year={2025}
}

@inproceedings{ModServe,
author = {Qiu, Haoran and Biswas, Anish and Zhao, Zihan and Mohan, Jayashree and Khare, Alind and Choukse, Esha and Goiri, \'{I}\~{n}igo and Zhang, Zeyu and Shen, Haiying and Bansal, Chetan and Ramjee, Ram and Fonseca, Rodrigo},
title = {ModServe: Modality- and Stage-Aware Resource Disaggregation for Scalable Multimodal Model Serving},
year = {2026},
isbn = {9798400722769},
address = {New York, USA},
doi = {10.1145/3772052.3772254},
booktitle = {Proceedings of the 2025 ACM Symposium on Cloud Computing},
pages = {817–830},
numpages = {14},
location = {
},
}

@inproceedings{DistServe,
author = {Zhong, Yinmin and Liu, Shengyu and Chen, Junda and Hu, Jianbo and et al.},
title = {DistServe: disaggregating prefill and decoding for goodput-optimized large language model serving},
year = {2024},
isbn = {978-1-939133-40-3},
publisher = {USENIX Association},
address = {USA},
booktitle = {Proceedings of the 18th USENIX Conference on Operating Systems Design and Implementation},
articleno = {11},
numpages = {18},
location = {CA, USA}
}

@book{ARIMA,
author = {Box, George Edward Pelham and Jenkins, Gwilym},
title = {Time Series Analysis, Forecasting and Control},
year = {1990},
isbn = {0816211043},
publisher = {Holden-Day, Inc.},
address = {USA}
}

@INPROCEEDINGS{splitwise,
  author={Patel, Pratyush and Choukse, Esha and Zhang, Chaojie and Shah, Aashaka and Goiri, Íñigo and Maleki, Saeed and Bianchini, Ricardo},
  booktitle={2024 ACM/IEEE 51st Annual International Symposium on Computer Architecture}, 
  title={Splitwise: Efficient Generative LLM Inference Using Phase Splitting}, 
  year={2024},
  volume={},
  number={},
  pages={118-132},
  doi={10.1109/ISCA59077.2024.00019}}

@misc{pdserve,
      title={P/D-Serve: Serving Disaggregated Large Language Model at Scale}, 
      author={Yibo Jin and Tao Wang and Huimin Lin and Mingyang Song and Peiyang Li and Yipeng Ma and et al.},
      year={2024},
      eprint={2408.08147},
      archivePrefix={arXiv},
      primaryClass={cs.DC},
      url={https://arxiv.org/abs/2408.08147}, 
}

@inproceedings{vllm,
  title={Efficient Memory Management for Large Language Model Serving with PagedAttention},
  author={Woosuk Kwon and Zhuohan Li and Siyuan Zhuang and Ying Sheng and Lianmin Zheng and Cody Hao Yu and et al.},
  booktitle={Proceedings of the ACM SIGOPS 29th Symposium on Operating Systems Principles},
  year={2023}
}

@online{ai-dynamo,
  author  = {Nvidia},
  title   = {ai-dynamo},
  year    = {2025},
  url     = {https://github.com/ai-dynamo/dynamo},
  urldate = {2025-07-01},
  note = {Accessed: 2025-07-01}
}

@online{nixl,
  author  = {Nvidia},
  title   = {NIXL},
  year    = {2025},
  url     = {https://github.com/ai-dynamo/nixl},
  urldate = {2025-07-01},
  note = {Accessed: 2025-07-01}
}

@online{DeepSeek-R1-Distill-Llama-70B-FP8-dynamic,
  author       = {
RedHatAI},
  title        = {RedHatAI/DeepSeek-R1-Distill-Llama-70B-FP8-dynamic},
  year         = {2025},
  url     = {https://huggingface.co/RedHatAI/DeepSeek-R1-Distill-Llama-70B-FP8-dynamic},
  urldate = {2025-07-01},
  note = {Accessed: 2025-07-01}
}

@online{opt-30b,
  author       = {facebook},
  title        = {facebook/opt-30b},
  year         = {2022},
  url     = {https://huggingface.co/facebook/opt-30b},
  urldate = {2025-07-01},
  note = {Accessed: 2025-07-01}
}

@online{sharegpt,
  author       = {anon8231489123},
  title        = {ShareGPT},
  year         = {2023},
  url     = {https://huggingface.co/datasets/anon8231489123/ShareGPT_Vicuna_unfiltered},
  urldate = {2025-07-01},
  note = {Accessed: 2025-07-01}
}

@online{AzurePublicDataset,
  author       = {Microsoft},
  title        = {AzurePublicDataset},
  year         = {2023},
  url     = {https://github.com/Azure/AzurePublicDataset},
  urldate = {2025-07-01},
  note = {Accessed: 2025-07-01}
}

@misc{burstgpt,
      title={BurstGPT: A Real-world Workload Dataset to Optimize LLM Serving Systems}, 
      author={Yuxin Wang and Yuhan Chen and Zeyu Li and Xueze Kang and Yuchu Fang and Yeju Zhou and et al.},
      year={2025},
      eprint={2401.17644},
      archivePrefix={arXiv},
      primaryClass={cs.DC},
      url={https://arxiv.org/abs/2401.17644}, 
}

@inproceedings{sglang,
author = {Zheng, Lianmin and Yin, Liangsheng and Xie, Zhiqiang and Sun, Chuyue and Huang, Jeff and Yu, Cody and et al.},
title = {SGLang: efficient execution of structured language model programs},
year = {2024},
isbn = {9798331314385},
publisher = {Curran Associates Inc.},
address = {NY, USA},
booktitle = {Proceedings of the 38th International Conference on Neural Information Processing Systems},
articleno = {2000},
numpages = {27},
location = {Vancouver, BC, Canada}
}

@inproceedings {Mooncake,
author = {Ruoyu Qin and Zheming Li and Weiran He and Jialei Cui and Feng Ren and Mingxing Zhang and et al.},
title = {Mooncake: Trading More Storage for Less Computation {\textemdash} A {KVCache-centric} Architecture for Serving {LLM} Chatbot},
booktitle = {23rd USENIX Conference on File and Storage Technologies},
year = {2025},
isbn = {978-1-939133-45-8},
address = {Santa Clara, CA},
pages = {155--170},
publisher = {USENIX Association},
month = feb
}

@misc{arrow,
      title={Arrow: Adaptive Scheduling Mechanisms for Disaggregated LLM Inference Architecture}, 
      author={Yu Wu and Tongxuan Liu and Yuting Zeng and Siyu Wu and Jun Xiong and Xianzhe Dong and et al.},
      year={2025},
      eprint={2505.11916},
      archivePrefix={arXiv},
      primaryClass={cs.DC},
      url={https://arxiv.org/abs/2505.11916}, 
}

@online{gpt-5,
  author       = {openai},
  title        = {Introducing GPT-5},
  year         = {2025},
  url     = {https://openai.com/index/introducing-gpt-5},
  urldate = {2025-08-27},
  note = {Accessed: 2025-07-01}
}

@online{gpt-3.5,
  author       = {openai},
  title        = {Introducing APIs for GPT-3.5 Turbo and Whisper},
  year         = {2024},
  url     = {https://openai.com/index/introducing-chatgpt-and-whisper-apis},
  urldate = {2025-08-27},
  note = {Accessed: 2025-07-01}
}

@inproceedings{transformer,
author = {Vaswani, Ashish and Shazeer, Noam and Parmar, Niki and Uszkoreit, Jakob and Jones, Llion and Gomez, Aidan N. and et al.},
title = {Attention is all you need},
year = {2017},
isbn = {9781510860964},
publisher = {Curran Associates Inc.},
address = {NY, USA},
booktitle = {Proceedings of the 31st International Conference on Neural Information Processing Systems},
pages = {6000–6010},
numpages = {11},
location = {Long Beach, California, USA}
}

@misc{kimi-k2,
      title={Kimi K2: Open Agentic Intelligence}, 
      author={Kimi Team and Yifan Bai and Yiping Bao and Guanduo Chen and Jiahao Chen and Ningxin Chen and et al.},
      year={2025},
      eprint={2507.20534},
      archivePrefix={arXiv},
      primaryClass={cs.LG},
      url={https://arxiv.org/abs/2507.20534}, 
}

@misc{llama2,
      title={Llama 2: Open Foundation and Fine-Tuned Chat Models}, 
      author={Hugo Touvron and Louis Martin and Kevin Stone and Peter Albert and Amjad Almahairi and Yasmine Babaei and et al.},
      year={2023},
      eprint={2307.09288},
      archivePrefix={arXiv},
      primaryClass={cs.CL},
      url={https://arxiv.org/abs/2307.09288}, 
}

@article{2024qwen2,
  title={Qwen2. 5-Coder Technical Report},
  author={Hui, Binyuan and Yang, Jian and Cui, Zeyu and Yang, Jiaxi and Liu, Dayiheng and Zhang, Lei and et al.},
  journal={arXiv preprint arXiv:2409.12186},
  year={2024}
}

@online{gemini,
  author       = {Google},
  title        = {Google Gemini},
  year         = {2025},
  url     = {https://blog.google/products/gemini},
  urldate = {2025-08-27},
  note = {Accessed: 2025-07-01}
}

@online{Cursor,
  author       = {Cursor},
  title        = {Cursor},
  year         = {2025},
  url     = {https://cursor.com},
  urldate = {2025-08-27},
  note = {Accessed: 2025-07-01}
}

@online{New-Bing,
  author       = {Jordi Ribas},
  title        = {New Bing},
  year         = {2023},
  url     = {https://blogs.bing.com/search-quality-insights/february-2023/Building-the-New-Bing},
  urldate = {2025-08-27},
  note = {Accessed: 2025-07-01}
}

@online{prometheus,
  author       = {Jordi Ribas},
  title        = {prometheus},
  year         = {2015},
  url     = {https://github.com/prometheus/prometheus},
  urldate = {2025-07-1},
  note = {Accessed: 2025-07-01}
}

@online{Llama-4,
  author       = {meta},
  title        = {The Llama 4 herd: The beginning of a new era of natively multimodal AI innovation},
  year         = {2025},
  url     = {https://ai.meta.com/blog/llama-4-multimodal-intelligence},
  urldate = {2025-08-27},
  note = {Accessed: 2025-07-01}
}

@online{H100,
  author       = {Nvidia},
  title        = {NVIDIA H100 Tensor Core GPU},
  year         = {2022},
  url     = {https://www.nvidia.com/en-us/data-center/h100},
  accessed = {2025-08-27},
  note = {Accessed: 2025-07-01}
}

@online{nvml,
  author       = {Nvidia},
  title        = {NVIDIA Management Library (NVML)},
  year         = {2025},
  url     = {https://developer.nvidia.com/management-library-nvml},
  urldate = {2025-08-27},
  note = {Accessed: 2025-07-01}
}

@inproceedings{flashattention1,
 author = {Dao, Tri and Fu, Dan and Ermon, Stefano and Rudra, Atri and R\'{e}, Christopher},
 booktitle = {Advances in Neural Information Processing Systems},
 pages = {16344--16359},
 publisher = {Curran Associates, Inc.},
 title = {FlashAttention: Fast and Memory-Efficient Exact Attention with IO-Awareness},
 volume = {35},
 year = {2022}
}

@ARTICLE{MoE,
  author={Jacobs, Robert A. and Jordan, Michael I. and Nowlan, Steven J. and et al.},
  journal={Neural Computation}, 
  title={Adaptive Mixtures of Local Experts}, 
  year={1991},
  volume={3},
  number={1},
  pages={79-87},
  doi={10.1162/neco.1991.3.1.79}}

@inproceedings {taming,
author = {Amey Agrawal and Nitin Kedia and Ashish Panwar and Jayashree Mohan and Nipun Kwatra and Bhargav Gulavani and et al.},
title = {Taming {Throughput-Latency} Tradeoff in {LLM} Inference with {Sarathi-Serve}},
booktitle = {18th USENIX Symposium on Operating Systems Design and Implementation},
year = {2024},
isbn = {978-1-939133-40-3},
address = {Santa Clara, CA},
pages = {117--134},
publisher = {USENIX Association},
month = jul
}

@misc{variable_length,
      title={LLM Serving Optimization with Variable Prefill and Decode Lengths}, 
      author={Meixuan Wang and Yinyu Ye and Zijie Zhou},
      year={2025},
      eprint={2508.06133},
      archivePrefix={arXiv},
      primaryClass={math.OC},
      url={https://arxiv.org/abs/2508.06133}, 
}

@article{ShuffleInfer,
author = {Hu, Cunchen and Huang, Heyang and Xu, Liangliang and Chen, Xusheng and Wang, Chenxi and Xu, Jiang and et al.},
title = {ShuffleInfer: Disaggregate LLM Inference for Mixed Downstream Workloads},
year = {2025},
issue_date = {June 2025},
publisher = {Association for Computing Machinery},
address = {New York, NY, USA},
volume = {22},
number = {2},
issn = {1544-3566},
url = {https://doi.org/10.1145/3732941},
doi = {10.1145/3732941},
journal = {ACM Trans. Archit. Code Optim.},
month = jul,
articleno = {77},
numpages = {24}
}

@inbook{PODAttention,
author = {Kamath, Aditya K. and Prabhu, Ramya and Mohan, Jayashree and Peter, Simon and et al.},
title = {POD-Attention: Unlocking Full Prefill-Decode Overlap for Faster LLM Inference},
year = {2025},
isbn = {9798400710797},
publisher = {Association for Computing Machinery},
address = {NY, USA},
booktitle = {Proceedings of the 30th ACM International Conference on Architectural Support for Programming Languages and Operating Systems, Volume 2},
pages = {897–912},
numpages = {16}
}

@ARTICLE{LLM4Edge,
author={Yao, Zhi and Tang, Zhiqing and Yang, Wenmian and Jia, Weijia},
journal={ IEEE Transactions on Services Computing },
title={{ Enhancing LLM QoS Through Cloud-Edge Collaboration: A Diffusion-Based Multi-Agent Reinforcement Learning Approach }},
year={2025},
volume={18},
number={03},
ISSN={1939-1374},
pages={1412-1427},
doi={10.1109/TSC.2025.3562362},
publisher={IEEE Computer Society},
address={Los Alamitos, CA, USA},
month=may}

@ARTICLE{LLM4Mobile,
  author={Li, Zonghang and Feng, Wenjiao and Guizani, Mohsen and Yu, Hongfang},
  journal={IEEE Transactions on Services Computing}, 
  title={TPI-LLM: Serving 70B-scale LLMs Efficiently on Low-resource Mobile Devices}, 
  year={2025},
  volume={},
  number={},
  pages={1-13},
  doi={10.1109/TSC.2025.3596892}}

@ARTICLE{LLM4EdgeScheduling,
author={Li, Yandi and Guo, Jianxiong and Tang, Zhiqing and Ding, Xingjian and Wang, Juncheng and et al.},
journal={ IEEE Transactions on Services Computing },
title={{ Cloud-Edge System for Scheduling Unpredictable LLM Requests with Combinatorial Bandit }},
year={2025},
volume={},
number={01},
ISSN={1939-1374},
pages={1-15},
doi={10.1109/TSC.2025.3611379},
publisher={IEEE Computer Society},
address={Los Alamitos, CA, USA},
month=sep}

@ARTICLE{LLM4Diagnosis,
author={Sun, Yongqian and Ma, Shiyu and Xiao, Tong and Zhao, Yongxin and Cai, Xuhui and Dong, Wei and et al.},
journal={ IEEE Transactions on Services Computing },
title={{ Accurate and Interpretable Log-Based Fault Diagnosis using Large Language Models }},
year={2025},
volume={},
number={01},
ISSN={1939-1374},
pages={1-14},
doi={10.1109/TSC.2025.3599494},
publisher={IEEE Computer Society},
address={Los Alamitos, CA, USA},
month=aug}

@inproceedings {Orca,
author = {Gyeong-In Yu and Joo Seong Jeong and Geon-Woo Kim and Soojeong Kim and Byung-Gon Chun},
title = {Orca: A Distributed Serving System for {Transformer-Based} Generative Models},
booktitle = {16th USENIX Symposium on Operating Systems Design and Implementation},
year = {2022},
isbn = {978-1-939133-28-1},
address = {Carlsbad, CA},
pages = {521--538},
publisher = {USENIX Association},
month = jul
}

@inproceedings{NSA,
    title = "Native Sparse Attention: Hardware-Aligned and Natively Trainable Sparse Attention",
    author = "Yuan, Jingyang  and
      Gao, Huazuo  and
      Dai, Damai  and
      Luo, Junyu  and
      Zhao, Liang  and
      Zhang, Zhengyan and et al.",
    booktitle = "Proceedings of the 63rd Annual Meeting of the Association for Computational Linguistics (Volume 1: Long Papers)",
    month = jul,
    year = "2025",
    address = "Vienna, Austria",
    publisher = "Association for Computational Linguistics",
    doi = "10.18653/v1/2025.acl-long.1126",
    pages = "23078--23097",
    ISBN = "979-8-89176-251-0"
}
\begin{IEEEbiography}[{\includegraphics[width=1in,height=1.25in,clip,keepaspectratio]{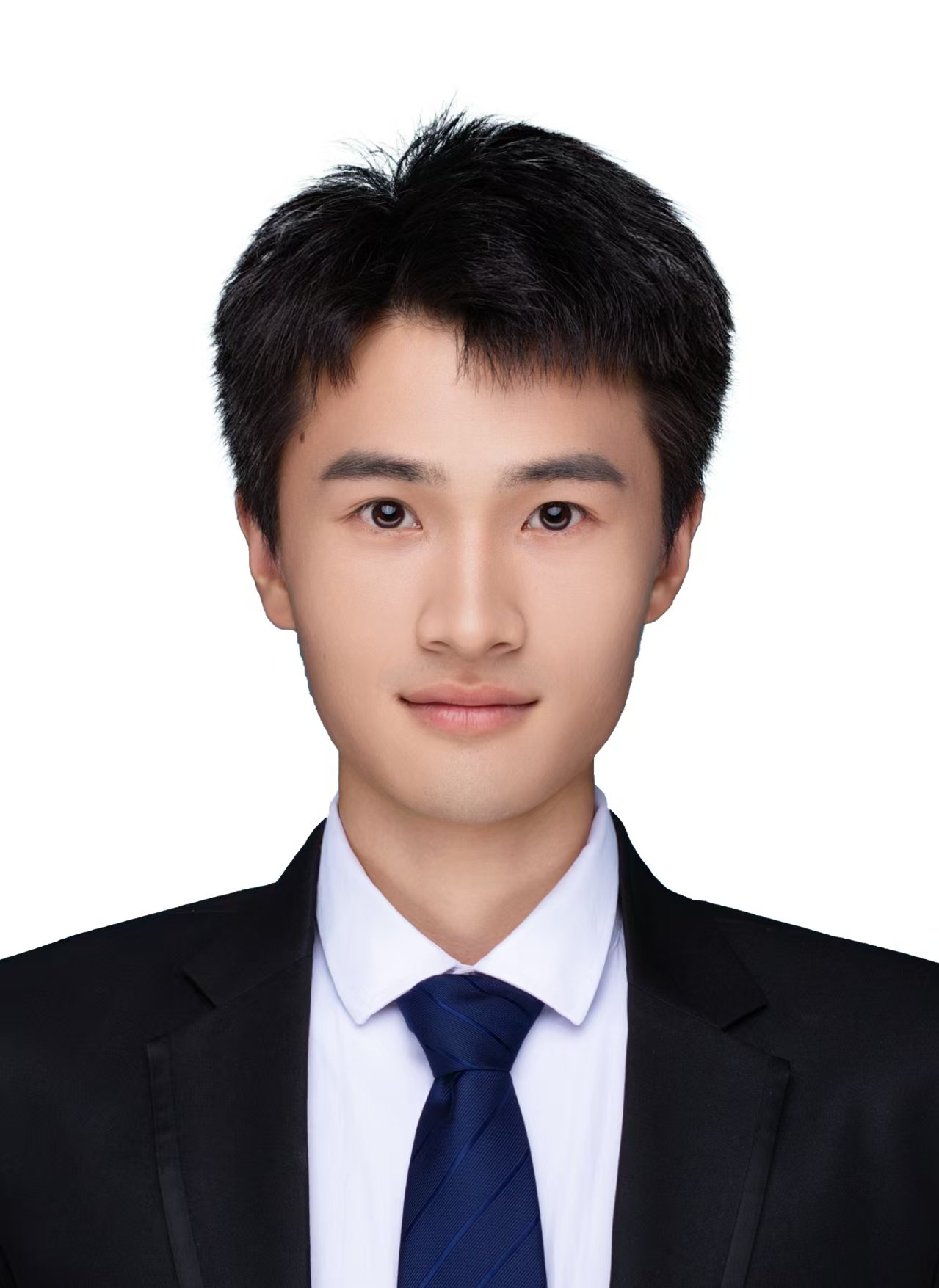}}]{Junhan Liao}
received his bachelor's degree in computer science and technology from Hunan University of Technology. Currently, he is a master's student at Shenzhen Institutes of Advanced Technology, Chinese Academy of Sciences. His main research interests include large language models inference optimization and system management.

\end{IEEEbiography}

\begin{IEEEbiography}[{\includegraphics[width=1in,height=1.25in,clip,keepaspectratio]{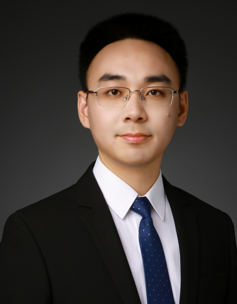}}]{Minxian Xu}

(Senior Member, IEEE) is currently an Associate Professor at Shenzhen Institutes of Advanced Technology, Chinese Academy of Sciences. He received his PhD degree from the University of Melbourne in 2019. His research interests include resource scheduling and optimization in cloud computing. He has co-authored 80+ peer-reviewed papers published in prominent international journals and conferences.
His PhD thesis was awarded the 2019 IEEE TCSC Outstanding Ph.D. Dissertation Award. He was also awarded the 2023 IEEE TCSC Award for Excellence (Early Career Award).

\end{IEEEbiography}

\begin{IEEEbiography}[{\includegraphics[width=1in,height=1.25in,clip,keepaspectratio]{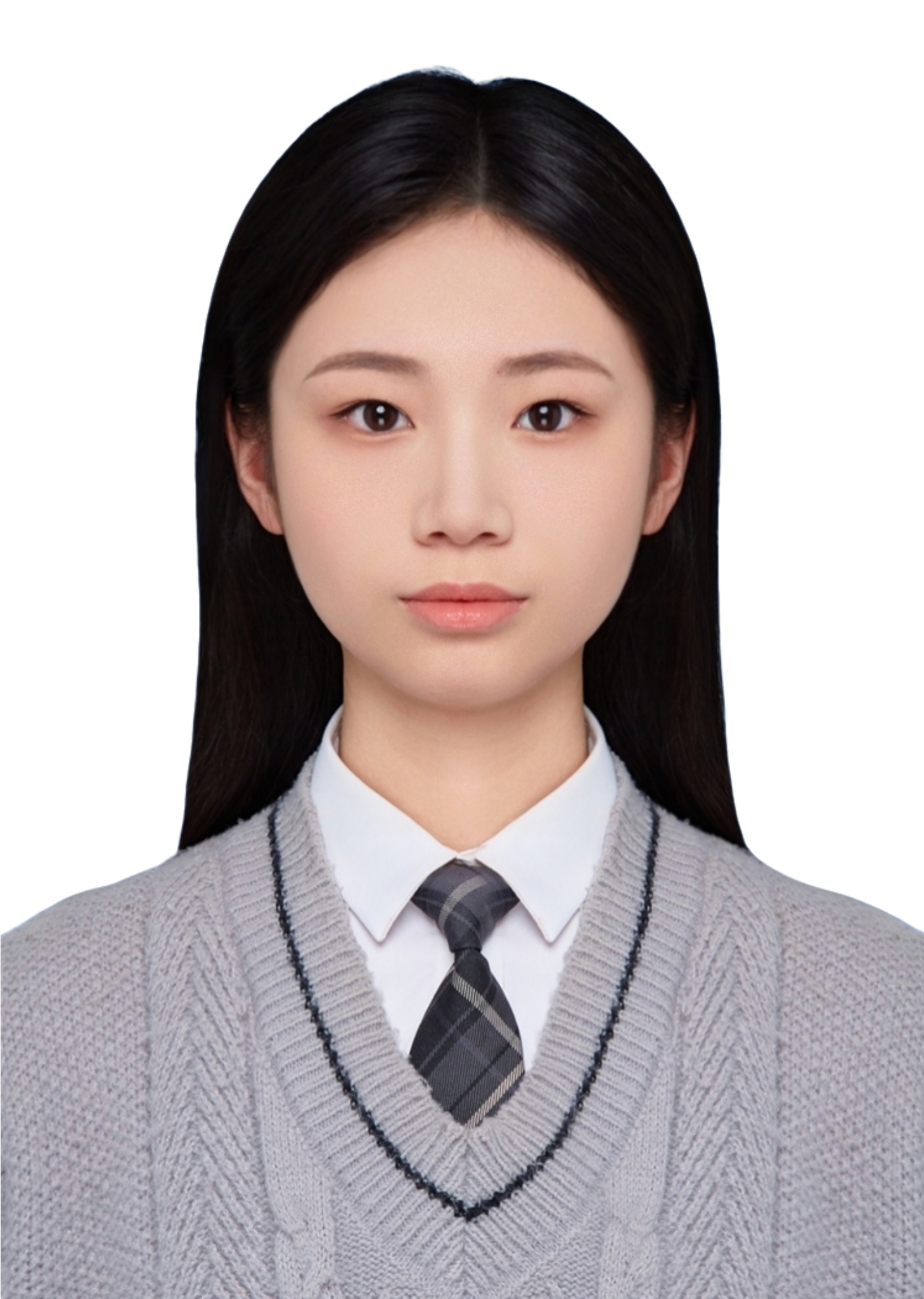}}]{Wanyi Zheng}
 received her Bachelor's degree from Hebei University in 2025. She is currently pursuing a Master's degree at the Southern University of Science and Technology, and is also a joint-training student at the Shenzhen Institutes of Advanced Technology, Chinese Academy of Sciences. Her research interests include operating systems and the integration of large language model systems, with a primary focus on LLM inference optimization and system management.

\end{IEEEbiography}
\begin{IEEEbiography}[{\includegraphics[width=1in,height=1.25in,clip,keepaspectratio]{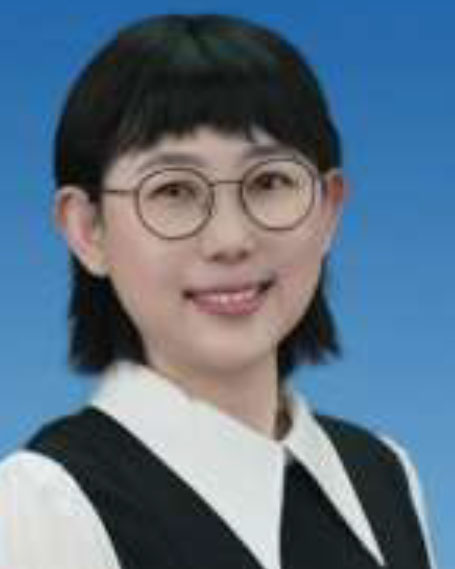}}]{Yan Wang}
(Member, IEEE) received the PhD degree in computer science from Inner Mongolia University, Hohhot, China, in 2015. She is currently an Associate Professor and a Master Supervisor of Computer Science and Technology with Inner Mongolia University. She has published more than 40 papers in international conferences and journals in her field. Her research interests include service computing, formal methods, and software technology.
\end{IEEEbiography}
\begin{IEEEbiography}[{\includegraphics[width=1in,height=1.25in,clip,keepaspectratio]{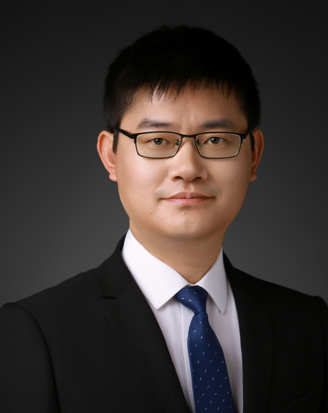}}]{Kejiang Ye}
(Senior Member, IEEE) received the BSc and PhD degrees from Zhejiang University, in 2008 and 2013, respectively. He was also a joint PhD student with the University of Sydney from 2012 to 2013. After graduation, he works as post-doc researcher with Carnegie Mellon University from 2014 to 2015 and Wayne State University from 2015 to 2016. He is currently a professor with the Shenzhen Institutes of Advanced Technology, Chinese Academy of Sciences. His research interests focus on the performance, energy, and reliability of cloud computing and network systems.
\end{IEEEbiography}

\begin{IEEEbiography}[{\includegraphics[width=1in,height=1.25in,clip,keepaspectratio]{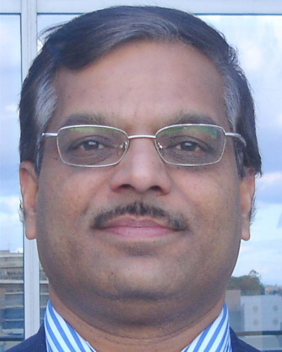}}]{Rajkumar Buyya}
(Fellow, IEEE) is currently a Redmond Barry distinguished professor and director with the Quantum Cloud Computing and Distributed Systems (qCLOUDS) Laboratory, the University of Melbourne, Australia. He has authored more than 625 publications and seven textbooks including ”Mastering Cloud Computing” published by McGraw Hill, China Machine Press, and Morgan Kaufmann for Indian, Chinese and international markets, respectively. He is one of the highly cited authors in computer science and software engineering worldwide (h-index=175, i10-index=837, 164,000+citations).
\end{IEEEbiography}
\begin{IEEEbiography}[{\includegraphics[width=1in,height=1.25in,clip,keepaspectratio]{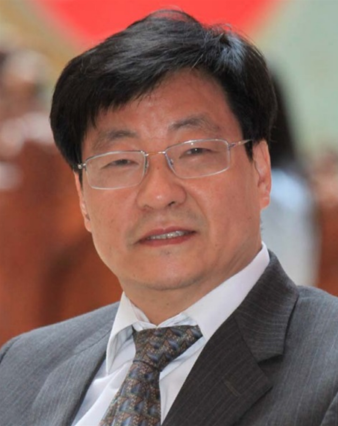}}]{Chengzhong Xu}
(Fellow, IEEE) received the Ph.D. degree in computer science and engineering from the University of Hong Kong in 1993. He is the Dean of Faculty of Science and Technology and the Interim Director of Institute of Collaborative Innovation, University of Macau. He published two research monographs and more than 300 peer-reviewed papers in journals and conference proceedings. His papers received about 17K citations with an H-index
of 72. His main research interests lie in parallel and
distributed computing and cloud computing.
\end{IEEEbiography}
\vfill

\end{document}